\documentclass[a4paper,twocolumn,11pt, unpublished]{quantumarticle}
\pdfoutput=1
\usepackage[utf8]{inputenc}
\usepackage[english]{babel}
\usepackage[T1]{fontenc}
\usepackage{amsmath}
\usepackage{hyperref}

\usepackage{tikz}
\usepackage{lipsum}

\usepackage{graphicx}
\usepackage{dcolumn}
\usepackage{bm}
\usepackage[utf8]{inputenc}
\usepackage[T1]{fontenc}
\usepackage{subfigure}
\usepackage{graphicx}
\usepackage{color}
\usepackage{enumerate}
\usepackage{footnote}
\usepackage{tikz} 
\usepackage{xcolor}
\usepackage{lineno}
\usepackage[numbers]{natbib}

\usepackage{braket}
\usepackage{multirow}
\newcommand*\emptycirc[1][1ex]{\tikz\draw (0,0) circle (#1);}

\newcommand*\fullcirc[1][1ex]{\tikz\fill (0,0) circle (#1);}

\begin{document}

\title{An Unbiased Quantum Random Number Generator Based on Boson Sampling}

\author{Jinjing Shi}
\affiliation{School of Computer Science and Engineering, Central South University, Changsha 410083, China.}
\email{shijinjing@csu.edu.cn.}
\author{Tongge Zhao}
\affiliation{School of Computer Science and Engineering, Central South University, Changsha 410083, China.}
\author{Yizhi Wang}
\affiliation{Institute for Quantum Information \& State Key Laboratory of High Performance Computing, College of Computer Science and Technology, National University of Defense Technology, Changsha 410073, China.}
\author{Chunlin Yu}
\affiliation{China Greatwall Quantum Laboratory, China Greatwall Technology Group CO.,LTD., Changsha 410073, China.}
\author{Yuhu Lu}
\affiliation{School of Computer Science and Engineering, Central South University, Changsha 410083, China.}
\author{Ronghua Shi}
\affiliation{School of Computer Science and Engineering, Central South University, Changsha 410083, China.}
\author{Shichao Zhang}
\affiliation{School of Computer Science and Engineering, Central South University, Changsha 410083, China.}
\email{zhangsc@mailbox.gxnu.edu.cn.}
\author{Junjie Wu}
\affiliation{Institute for Quantum Information \& State Key Laboratory of High Performance Computing, College of Computer Science and Technology, National University of Defense Technology, Changsha 410073, China.}
\email{junjiewu@nudt.edu.cn}

\maketitle

\begin{abstract}
  It has been proven that Boson sampling is a much promising model of optical quantum computation, which has been applied to designing quantum computer successfully, such as \emph{"Jiuzhang"}. However, the meaningful randomness of Boson sampling results, whose correctness and significance were proved from a specific quantum mechanical distribution, has not been utilized or exploited. In this research, Boson sampling is applied to design a novel Quantum Random Number Generator (QRNG) by fully exploiting the randomness of Boson sampling results, and its prototype system is constructed with the programmable silicon photonic processor, which can generate uniform and unbiased random sequences and overcome the shortcomings of the existing discrete QRNGs such as source-related, high demand for the photon number resolution capability of the detector and slow self-detection generator speed. Boson sampling is implemented as a random entropy source, and random bit strings with satisfactory randomness and uniformity can be obtained after post-processing the sampling results. It is the first approach for applying the randomness of Boson sampling results to develop a practical prototype system for actual tasks, and the experiment results demonstrate the designed Boson sampling-based QRNG prototype system pass 15 tests of the NIST SP 800-22 statistical test component, which prove that Boson sampling has great potential for practical applications with desirable performance besides quantum advantage.
\end{abstract}

\section{\label{sec:level1}Introduction}

Random numbers play an important role in different kinds of applications and provide essential resources \cite{caflisch1998monte}, such as cryptography \cite{shannon1949communication}, lottery industry \cite{chelkowski2010visualizing}, genetics, scientific simulations \cite{milz2021quantum, king2021scaling}, etc. While the generation of good randomness is an obviously difficult problem \cite{shakhovoy2020quantum,dale2015provable}. In the classical field, there are many random number generator (RNG) schemes, most of which involve deterministic algorithms and start from a small bit string called \emph{seed}, such as the linear congruential RNG based on number theory \cite{scherzer2010handbook}, the Mersenne Twister based on linear shift feedback registers \cite{matsumoto1998mersenne}, etc.
Those schemes can simply generate random numbers at an advantageous effective rate,
while they cannot be considered as true RNGs, called pseudorandom number generators (PRNGs). When the small string random \emph{seed} is leaked, the random numbers generated by those schemes may be exactly known. The important requirement for the unpredictability of random number generators limits the application scope of classical PRNGs.

Fortunately, the intrinsic randomness of quantum physics makes quantum systems a good source of RNG \cite{stefanov2000optical}, and many quantum random number generators (QRNGs) \cite{herrero2017quantum} are proposed, including Branching path generator \cite{jennewein2000fast}, Photon counting generator \cite{ren2011quantum}, etc. Branching path generator generates random numbers by measuring the path superposition state or polarization superposition state \cite{jennewein2000fast}, which depends on the input source that once the superposition state is controlled by a third-party, the measurement result may be controlled as well.
Photon counting generator obtains the random number by detecting the photon number of the coherent states \cite{ren2011quantum}, whose efficiency depends on the dead time and the photon resolving capability of photon detectors.
The non-local correlation between two particles can also generate the randomness \cite{pironio2010random} which provides an important resource in RNG protocols, like randomness expansion \cite{colbeck2011private, coudron2014infinite} and random amplification \cite{xu2017explicit}. These methods for constructing QRNGs are called device independent or self-testing QRNGs \cite{ma2016quantum, cao2016source, lunghi2015self}, while their generation bit rate is relatively slow that they lack wide applications.
In general, discrete QRNGs still have the following possible problems: source-related, high demand for the photon number resolution capability of the detector and slow self-detection generator speed. Thus the principle and implementation of QRNGs need to be improved urgently, and more and more new methods are expectively proposed to design QRNGs \cite{acerbi2020structures, zheng20196, prokhodtsov2020silicon}.

On the other hand, at the very beginning of noisy intermediate-scale quantum (NISQ) era \cite{preskill2018quantum}, quantum supremacy \cite{2016Characterizing} has been proposed that certain computing tasks may be executed exponentially faster on quantum computers than on classical computers\cite{harrow2017quantum}. Boson sampling is considered as a powerful candidate for experimental verification of quantum supremacy \cite{aaronson2013computational}, which samples the probability distribution generated by the evolution of single photons through a linear optical network only including phase shifters and beam splitters. The application of Boson sampling has attracted the attention of researchers, e.g. Boson sampling-based one-way function \cite{nikolopoulos2019cryptographic} , Boson sampling-based random unitary encryption \cite{feng2020arbitrated} and Boson sampling-based hash function \cite{shi2022quantum} have been proposed respectively. The branches of applications of Boson sampling shorten the research process of solving practical problems with Boson sampling. However, the meaningful randomness of Boson sampling results, whose correctness and significance were proved from a specific quantum mechanical distribution \cite{clifford2018classical}, has not been fully utilized or exploited.

We are so excited to discover that there are three favorable factors for taking advantages and exploiting the randomness of Boson sampling to design QRNG. First, the Boson sampling result is a superposition state before the photon output detection that the output is random and independent of the input source. Second, we merely need a normal single-photon detector instead of a photon number resolver in concluding the output of Boson sampling since we only concern whether a mode has photons instead of the specific number of photons. Third, it is possible to generate multi-bit random numbers for sampling only once with multi-modes Boson sampling, which may be more efficient compared to the generators for single-bit random numbers with one sampling. Moreover, the Boson sampling-based QRNG corresponding to above three factors is capable to overcome the defects of the existing QRNGs for the requirements of reliable safe input source, high photon number resolution and improved bit rate potentially. Boson sampling is a sampling problem which is closely related to random number generation indeed. As early as 1999, Petrie \textit{et al.} electronically amplified and sampled natural phenomena such as heat or shot noise signals to generate random numbers \cite{petrie1999sampling}. Zhou \textit{et al.} sampled vacuum fluctuations to design practical quantum random number generation \cite{zhou2019practical}. In 2021, Bai \textit{et al.} performed homodyne detection and sampling on the vacuum state to realize the fastest and miniaturized QRNG at present\cite{bai202118}.
In the field of experiments, many ingenious experiments have implemented small-scale Boson sampling
\cite{tillmann2013experimental, brod2019photonic}. At the same time, the Boson sampling model has also been extended, i.e. Scattershot Boson sampling \cite{bentivegna2015experimental}, Lossy Boson sampling \cite{wang2018toward, aaronson2016bosonsampling}, Gaussian Boson sampling \cite{hamilton2017gaussian, kruse2019detailed} etc. In 2019, Wang \textit{et al.} realized the 20-photons 60-modes Boson sampling experiment\cite{wang2019boson}. In 2020 and 2021, quantum computing prototypes \emph{jiuzhang} and \emph{jiuzhang 2.0} were proposed \cite{zhong2020quantum, zhu2021quantum}, respectively, which proved the advantages of quantum computing in the optical quantum system.

In order to overcome the above shortcomings of the existing QRNGs and in the interest of fully exploiting the randomness of Boson sampling results, we present an exploratory work in this paper to utilize Boson sampling as an entropy source to generate quantum random numbers, where the sampling results are independent with the input source, only a normal single-photon detector is required without photon number resolution, and the generator is quite efficient and effective with ideal post-processing. The Von Neumann correction method for post-processing is applied to generate uniform and unbiased random numbers. A Boson sampling-QRNG prototype system is constructed with the programmable silicon photonic processor to implement the experiments for NIST SP 800-22 statistical test \cite{bassham2010sp} of the proposed scheme.

\section{\label{sec:2}PRELIMINARIES}
In this section, we first briefly review the basic knowledge of Boson sampling, and then introduce the design methods of using quantum optical systems to construct QRNGs.

\subsection{\label{sec:2.1}Boson sampling model}

\begin{figure}[htbp]\label{F1}
\centering
\subfigure[]{
\includegraphics[width=8.4cm]{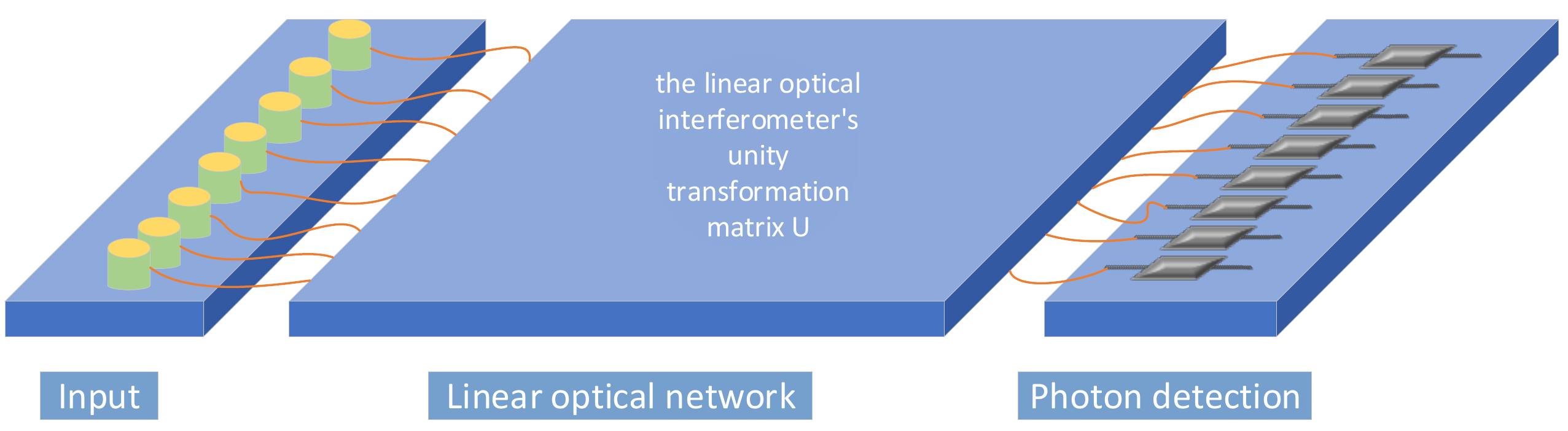}
\label{F1aa}
}
\subfigure[]{
\includegraphics[width=4cm]{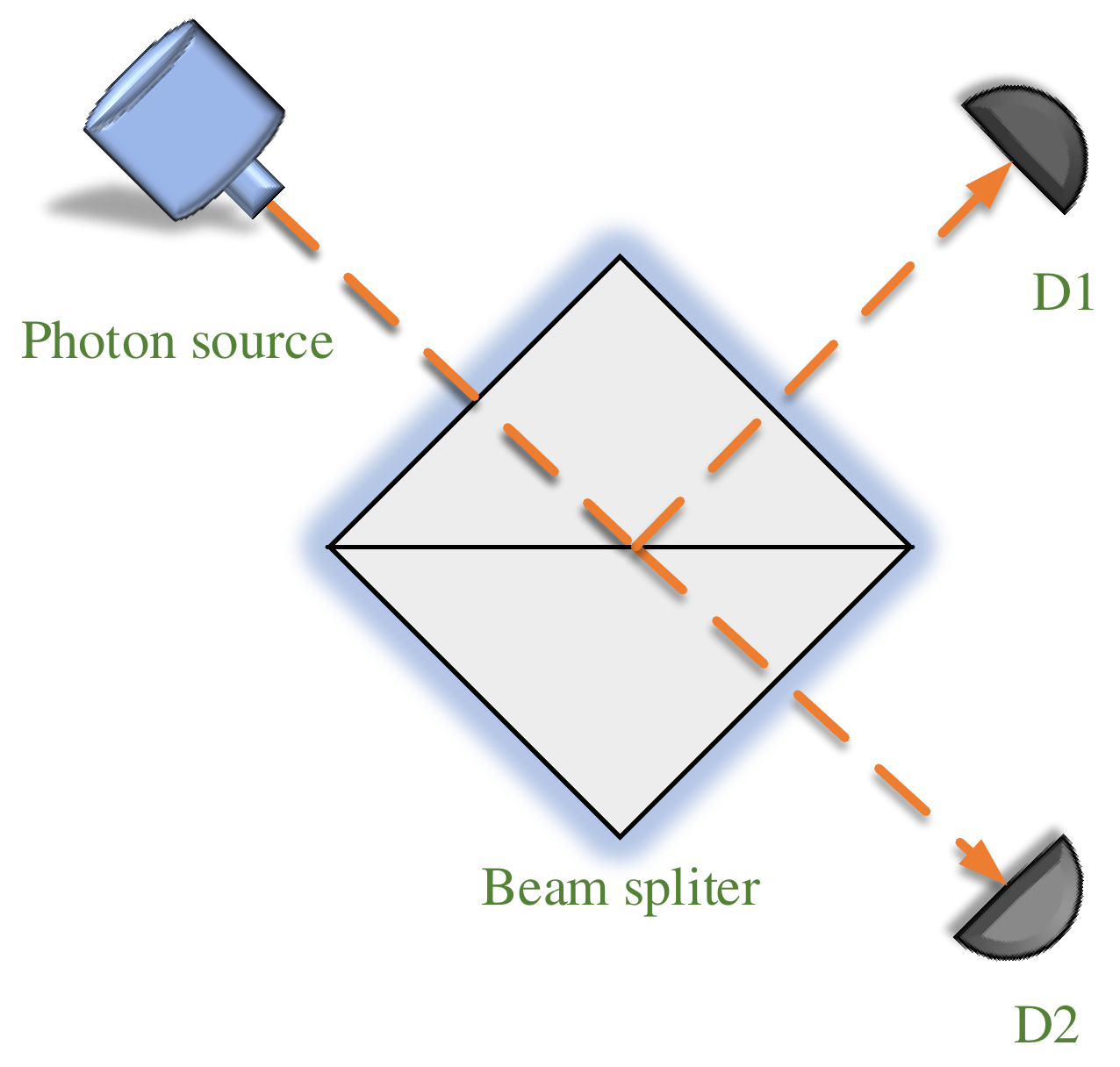}
\label{F1bb}
}
\caption{(a) $n$ photons $m$ modes Boson sampling model. $n$ Bosons are input into the passive linear network of $m$ modes, which is composed of beam splitters and phase shifters. After the interference effect of the network, the photons are detected and sampled by the detectors at the output modes. (b) The structure of Branching path QRNG. The photon is sent to a balanced beam splitter and the output is detected with the same probability.}
\end{figure}

The problem of Boson sampling is proposed by Aaronson and Arkhipov in 2013 \cite{aaronson2013computational}, claiming that the fock state is input into the passive linear optical interferometer, and the output distribution is sampled with the photon detectors, as shown in the Fig.\ref{F1aa}. Without the exponential cost of time or resources, the sampling outcome cannot be predicted with a classical computer \cite{aaronson2016complexity, lund2017exact}.
For Boson sampling model of $m$-modes linear transformation and $n$ indistinguishable Bosons, the output probability is calculated by a tricky matrix function called permanent
\begin{widetext}
\begin{equation}
  P_{I, O}=\left|\langle O|U \otimes U \ldots \otimes U| I\rangle\right|^{2}=\frac{\left|{Per}\left(U_{I, O}\right)\right|^{2}}{j_{1} ! j_{2} ! \ldots j_{m} ! g_{1} ! g_{2} ! \ldots g_{m} !},
\end{equation}
\end{widetext}
where $|I\rangle=\left|j_{1} j_{2} \ldots j_{m}\right\rangle$ is the input state satisfying $j_{1}+j_{2}+\ldots+j_{m}=n$, $j_{i}$  is the number of photons input to the $i$-th  mode, $|O\rangle=\left|g_{1} g_{2} \ldots g_{m}\right\rangle$ is the output state, $U_{I, O}$ is the sub-matrix of the unitary transformation matrix $U$ for the linear optical interferometer.

\subsection{\label{2.2} Quantum random number generator}

The uncertainty of quantum systems is guaranteed by the inherent randomness of quantum physics, where true randomness is a basic feature of quantum mechanics \cite{govcanin2022sample}. With the development of quantum optics, the inherent randomness in many parameters of the quantum states of light provides options for the realization of QRNGs \cite{herrero2017quantum}.

Branching path QRNGs generate random numbers based on the measurement of path superposition state. The light source is input into a balanced beam splitter, where the transmittance and reflectivity are equal. The output is 0 when $D_0$ detects a photon, and the output is 1 when $D_1$ detects a photon \cite{stefanov2000optical}, as shown in the Fig.\ref{F1bb}. Photon counting QRNGs generate random numbers based on the number of photons detected on the coherent state \cite{ren2011quantum}. Coherent states can be described as a superposition of fork states
\begin{equation}
\left|\alpha \right\rangle=e^{-\left | \alpha  \right |^{2}/2}\sum_{n=0}^{\infty }\frac{\alpha ^{n}}{\sqrt{n!}}\left|n \right\rangle,
\end{equation}
where $\alpha$ is a complex number, the amplitude $\left | \alpha  \right |^{2}$ corresponds to the mean photon number of the state. The number of photons arriving at the detector in a fixed time follows a Poisson distribution. The probability of detecting $n$ photons in $T$ time is
\begin{equation}
P\left ( n \right )=\frac{\left ( \lambda T \right )^{n}}{n!}e^{-\lambda T}.
\end{equation}

Boson sampling can be handled with optical quantum systems, of which the necessary elements can be implemented with the current technologies, i.e. sources, linear evolution and detection. The result of Boson sampling is random and can be used as a method of realizing a QRNG.

\section{\label{sec:3}Quantum random number generator based on Boson sampling}

In this section, we describe the overall scheme and individual modules of the Boson sampling-based QRNG. The Boson sampling-based QRNG prototype system is constructed on a programmable silicon photonic processor.

\subsection{\label{sec:3.1}The scheme of Boson sampling-based QRNG}

\begin{figure*}[htbp]\label{F-A}
\centering
\subfigure[]{
\includegraphics[width=12.5cm]{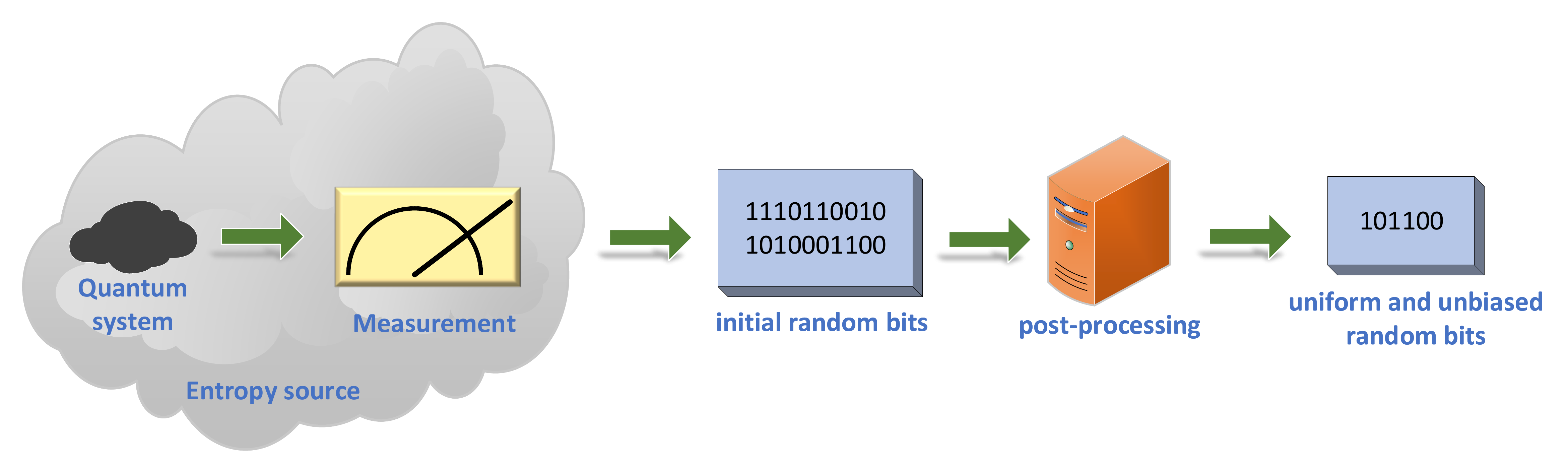}
\label{F1a}
}

\subfigure[]{
\includegraphics[width=12.5cm]{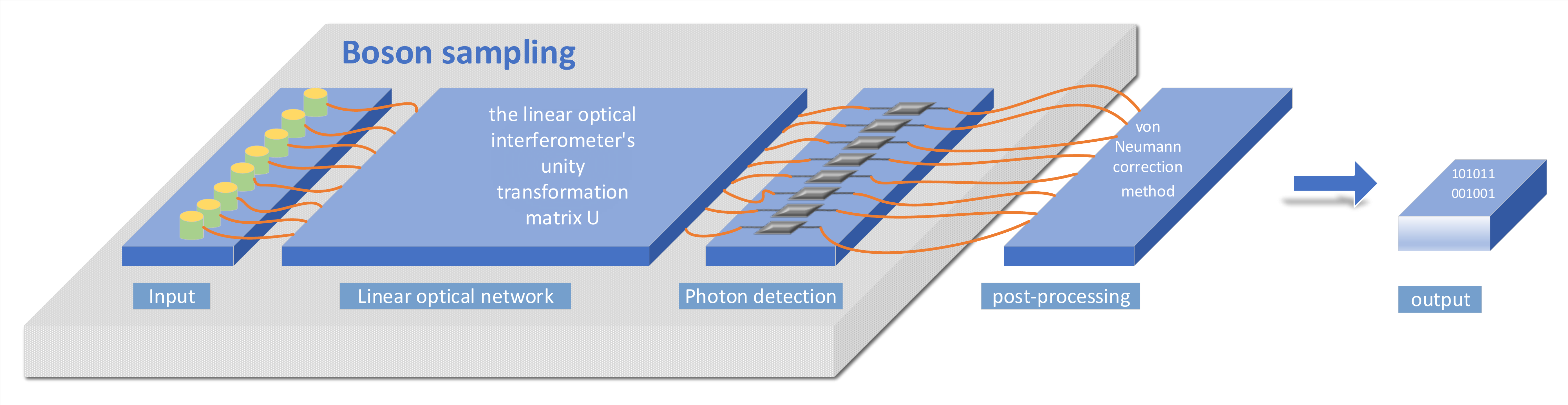}
\label{F1c}
}
\caption{(a) Block structure of QRNG. The block structure consists of the entropy source and the post-processing. The entropy source includes the quantum system and the measurement process, generating the initial random numbers. The post-processing ensures the random number sequence can be uniform and unbiased with 0bit and 1bit derived on equal probability. (b) The structure of Boson sampling-based QRNG. The sampling results are post-processed with Von Neumann correction method to obtain a uniform 01-bit sequence.}

\end{figure*}

The QRNGs are divided into well-defined independent modules according to their functions, as shown in Fig.\ref{F1a}, where the most important parts are the entropy source and the post-processing. The entropy source is composed of a quantum system and measurement operations to output initial random data. Since noise exists in the entire system and the measurement process, the initial random data is not necessarily in the form of random bits. Therefore, post-processing operations are required, such as encoding, random extraction, random amplification, etc, ensuring the random sequence obtained at the final output can become a uniform random number sequence with unbiased and independent bit 0 and bit 1.

\begin{table*}[tb]
  \begin{center}
      \begin{tabular}{c c c c c c c c c c c c c c c}
      \hline
      mode &1 & 2 & 3 & 4 & 5 & 6 & 7 & 8 & 9 & 10 &11 & 12 & ... &$M$ \\
      \hline
      The first sampling result $S_1$ &\fullcirc & \fullcirc & \emptycirc & \fullcirc & \emptycirc & \emptycirc & \fullcirc & \emptycirc & \fullcirc & \emptycirc &\emptycirc & \fullcirc & ...&\emptycirc \\
       \hline
      The second sampling result $S_2$ &\emptycirc & \fullcirc & \emptycirc & \emptycirc & \fullcirc & \emptycirc & \emptycirc & \fullcirc & \fullcirc & \fullcirc &\emptycirc & \emptycirc & ... &\fullcirc \\
      \hline
      Coding &0 & * & * & 0 & 1 & * & 0 & 1 & * & 1 &* & 0 & ... &1 \\
      \hline
      The final random number sequence& \multicolumn{14}{c}{0010110...1} \\
      \hline
      \end{tabular}
      \end{center}
      \footnotesize{$\fullcirc$ represents one or more photons is detected}\\
      \footnotesize{$\emptycirc$ represents no photon is detected}\\
   \caption{The post-processing with the Von Neumann correction method}
   \label{Tab1}
\end{table*}

In the Boson sampling-based QRNG scheme we designed, five major components are included as described in Fig.\ref{F1c}, i.e. input Boson source, linear optical network, photon detection, post-processing, and output. The entropy source is described by the Boson sampling, in which the quantum system is composed of the input Boson source and the linear optical network, and the measurement process is detected by the photon detectors.

With high probability, the Boson sampling distribution created by a haar-random matrix $U$ is far away from the uniform distribution\cite{aaronson2014bosonsampling}.
The Von Neumann correction method\cite{von195113} as shown in Table \ref{Tab1} is applied to the random numbers obtained directly from the Boson sampling to get a completely uniform and unbiased random binary bit sequence.
The specific steps of the post-processing are summarized as follows.

\begin{description}
\item [Step 1] Perform a Boson sampling and record the sampling result as $S_1$.
\item [Step 2] Perform Boson sampling again and record the result as $S_2$.
\item [Step 3]
Comparing $S_1$ with $S_2$, the output code of a mode is 0 bit if it detects photon in $S_1$ but does not detect photon in $S_2$. Similarly, a mode does not detect photon in $S_1$ but detects photon in $S_2$, then the output code of this mode is 1 bit. No coding in other cases.
\item [Step 4]
The coded bits of the modes numbered 1 to $M$ are output in order to obtain the result of the random number generator.
\item [Step 5]Repeat Step 1 to Step 4 until the required number of random numbers are generated
\end{description}

\begin{figure*}[tb]
\begin{center}
\includegraphics[scale=0.22]{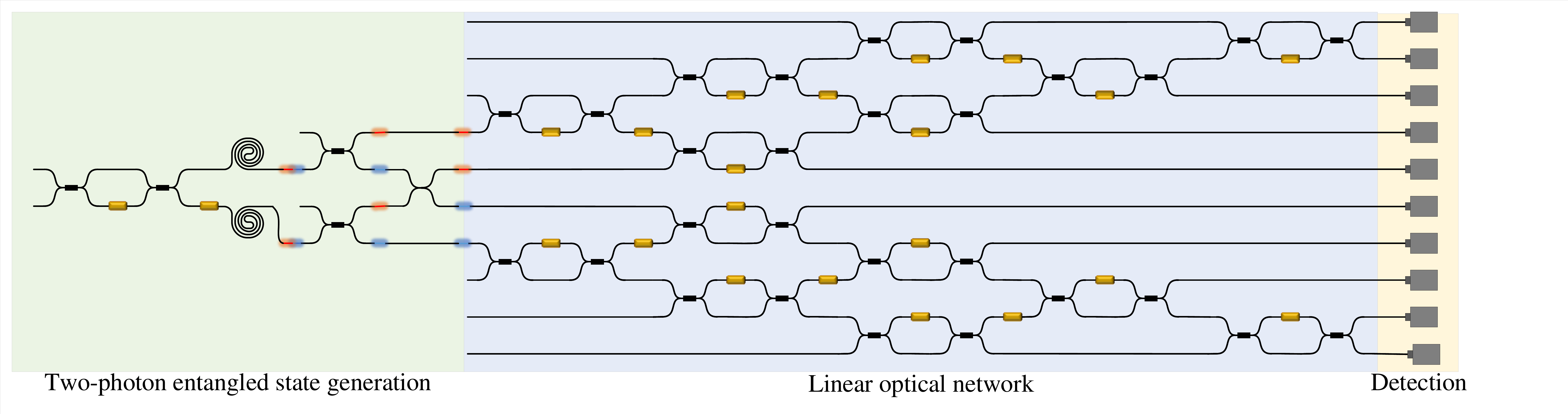}
\end{center}
\caption{The programmable silicon photonic processor chip. The silicon chip integrates two functional parts: (i) creating two-photon entangled state through spontaneous four-wave mixing (SFWM) and (ii) applying arbitrary five-dimensional unitary transformations with fixed inputs through configurable linear optical networks.}
\label{F2}
\end{figure*}

\begin{figure*}[tb]
\begin{center}
\subfigure[]{ \includegraphics[width=3.2in]{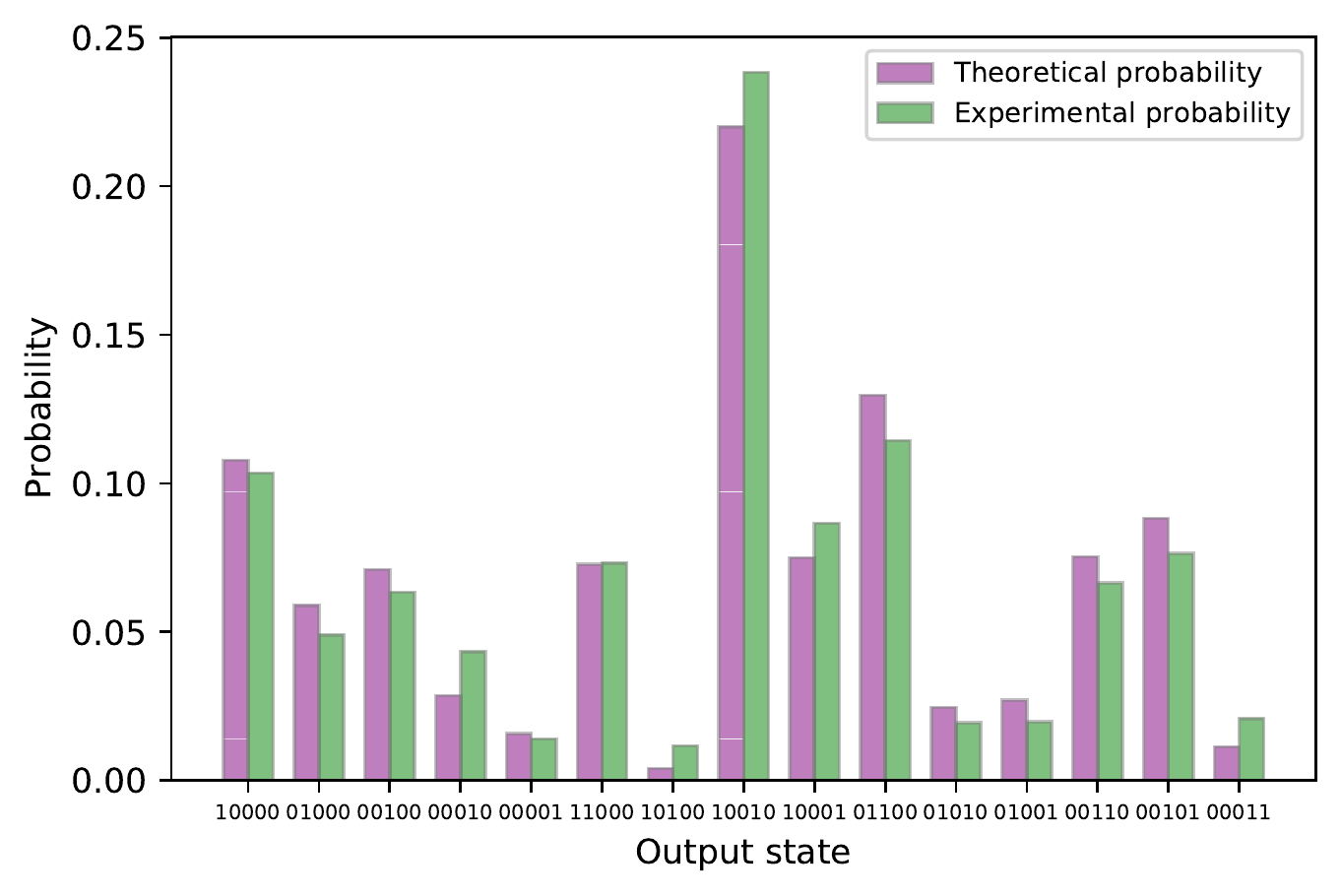}\label{F3a}}
\subfigure[]{ \includegraphics[width=4in]{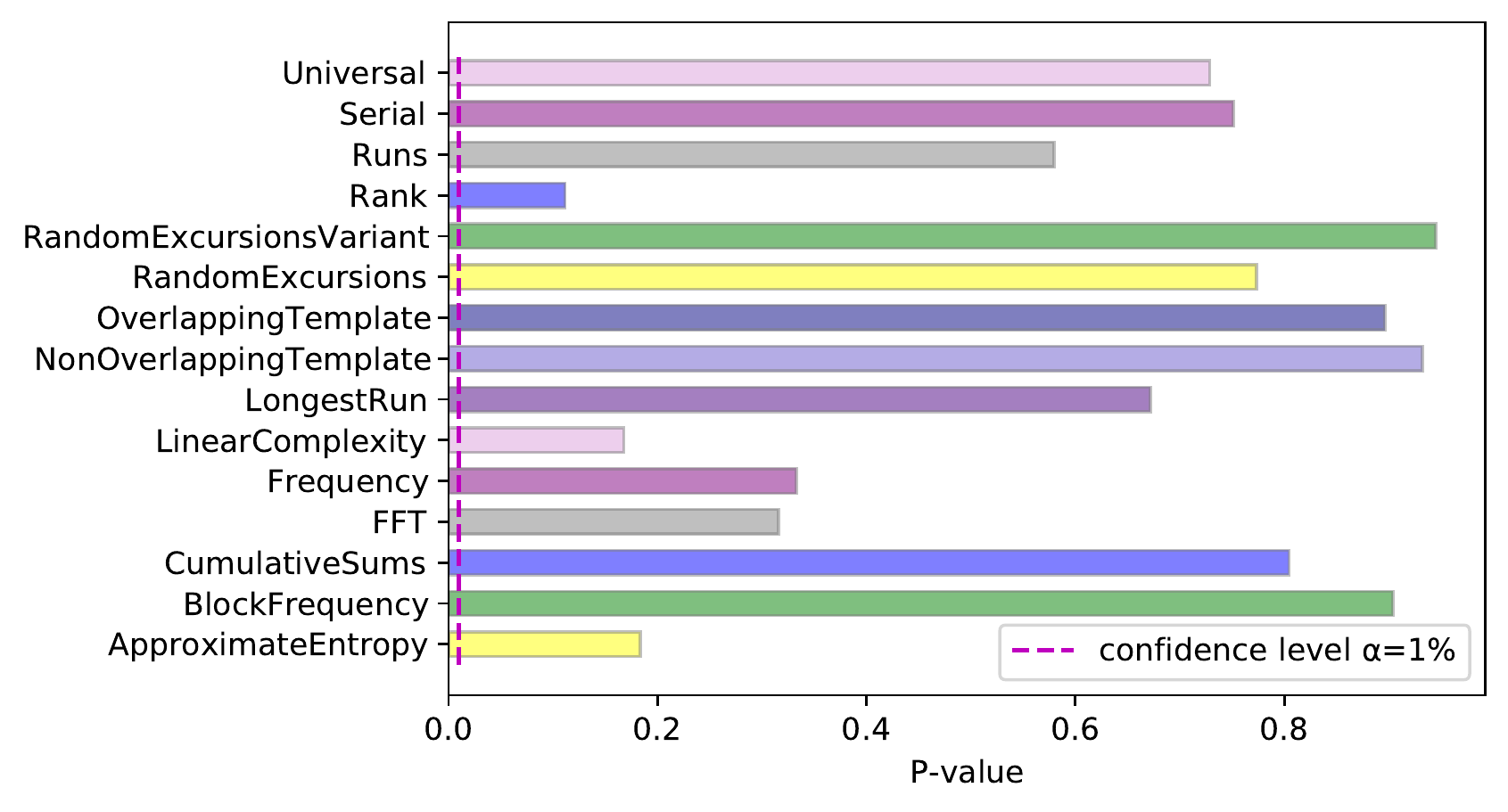}\label{F3c}}
\end{center}
\caption{(a) The theoretical probability distribution and experimental probability distribution of each output state. (b) The experimental results pass the 15 tests of the NIST SP 800-22 random number statistical tests. The $P-value$ of each test is greater than the confidence level $\alpha=0.01$.}
\end{figure*}

\subsection{\label{sec:3.22}The Boson sampling-based QRNG prototype system}

We have constructed the QRNG prototype system using a programmable silicon photonic processor \cite{qiang2021implementing} and observed two Bosons sampling statistics after a five-mode Haar random unitary process $U_5$, as shown schematically in Fig.\ref{F2}. To simulate a state of two Bosons created in modes $1$ and $2$, we first generate the two-photon entangled state
\begin{equation}\label{estate}
  \frac{1}{\sqrt{2}}\left(|1\rangle_{s}|2\rangle_{i}+|2\rangle_{s}|1\rangle_{i}\right),
\end{equation}
where $|k\rangle(k=1,2,3,4,5)$ denotes the path of a photon, $s$ and $i$ denote the signal and idler photon from coherently generated photon pairs. Each photon is then routed to a universal linear optical network, which applies the unitary transformation $U_5$ with fixed input. Finally, photons are detected off the chip, and two-photon coincidence events are recorded, from which we extract the correlated detection probability $P_{r,q}$ of measuring a photon at output $r$ and $q$ of two networks, respectively. According to the simulating scheme, we obtain  $ \Gamma_{r,q}$, the probability to measure two Bosons occupying modes $r$ and $q$ after the unitary process $U_5$, as $ \Gamma_{r,q}=P_{r,q}$ (shown in the Fig.\ref{F3a}).

\section{\label{sec:4}Experiments and Results}

In this subsection, we use the Boson sampling-based QRNG prototype system to experimentally generate random numbers.
Furthermore, the statistical test suite is performed on the random numbers obtained from experiments to test the randomness.
\begin{table*}
\centering
\begin{tabular}{|c||c|c|c|}
\hline
      number &Test  \\
      \hline
      1 &The Approximate Entropy Test  \\
      \hline
      2 &The Binary Matrix Rank Test \\
      \hline
      3 &The Cumulative Sums Test \\
      \hline
      4 &The Frequency (Monobit) Test  \\
      \hline
      5 &Frequency Test within a Block  \\
      \hline
      6 &The Runs Test  \\
      \hline
      7 &Tests for the Longest-Run-of-Ones in a Block  \\
      \hline
      8 &The Discrete Fourier Transform (Spectral) Test  \\
      \hline
      9 &The Non-overlapping Template Matching Test  \\
      \hline
      10 &The Overlapping Template Matching Test  \\
      \hline
      11 &Maurer's "Universal Statistical" Test  \\
      \hline
      12 &The Linear Complexity Test  \\
      \hline
      13 &The Serial Test  \\
      \hline
      14 &The Random Excursions Test  \\
      \hline
      15 &The Random Excursions Variant Test  \\
      \hline
\end{tabular}
\caption{The 15 tests of NIST SP 800-22}
\label{nist}
\end{table*}

\subsection{\label{sec:4.1}Experimental setup}

Bright light with the wavelength of 1549.3 nm is collected from a continuous-wave tunable laser and further amplified up to 50 mW using an optical erbium-doped fiber amplifier. After spectrally filtered by a dense wavelength division multiplexing module and tuned by a polarization controller, the bright light is then injected into the silicon photonic chip (shown in Fig.\ref{F2}) via a V-Groove fiber array. The chip mainly integrates two spontaneous four-wave mixing spiral sources, 22 simultaneously running thermo-optic phase shifters, 32 multimode interferometer beam splitters, and 16 optical grating couplers. The two SFWM sources are pumped by 50:50 split laser and create possible (signal-idler) maximally entangled photon pairs. The two five-mode universal linear optical networks comprised of beam splitters and phase shifters arranged in Reck \textit{et al.}–style are used to implement arbitrary five-mode unitary transformation with fixed inputs. In our prototype demonstration, we apply a Haar-random generated unitary operator $U_5$ as

\begin{widetext}

{\footnotesize
\begin{equation}
\begin{array}{lllll}
0.01311+0.33011*I & 0.65648+0.18793*I & 0.05465+0.16879*I & 0.49720+0.04980*I & 0.18845+0.32849*I \\
0.14759+0.47792*I & -0.24696-0.23840*I & 0.34173+0.03745*I & 0.25733+0.55151*I & -0.05311-0.37496*I \\
0.32357+0.04518*I & -0.10722+0.56648*I & 0.50920+0.17247*I & -0.29096-0.15241*I & 0.39168-0.10264*I \\
0.35818-0.46781*I & -0.13350-0.15289*I & -0.09973+0.00157*I & 0.07696+0.51280*I & 0.42931+0.38538*I \\
-0.31110+0.30001*I & -0.20562+0.00389*I & 0.13427-0.73030*I & -0.01605-0.05381*I & 0.35379+0.30204*I
\end{array}.
\end{equation}
}
\end{widetext}

Photons from different chip outputs are collected by the same fiber array and detected by superconducting nanowire single-photon detectors. With the coincidence counting logic, two-photon coincidences are recorded by using a time-interval-analyzer in a 468.75ps integration window. During the continuous running time, a total of 914286 two-photon coincidence events are recorded.

\subsection{\label{sec:4.2}Standard statistical randomness test}

A total of 1.8772 Mbit 01bit sequences were obtained after post-processing using the Von Neumann correction method, where the probabilities of 0 and 1bit are 0.49999814 and 0.50000186, respectively, implementing an unbiased distribution of random number sequences.
We use the NIST SP 800-22 statistical test suite to test the randomness \cite{bassham2010sp}, which is based on hypothesis testing, containing 15 tests in total, as shown in Table \ref{nist}.
A \emph{p-value} is calculated for each test item summarizing the degree of evidence rejecting the null hypothesis, which is the function of data. Choosing the significance level $\alpha=0.01$, if the \emph{p-value} obtained by the statistical test is greater than or equal to 0.01, it indicates that the sequence has passed the test, and the resulting sequence is random. If the \emph{p-value} is less than 0.01, it indicates that the sequence has failed the test. The \emph{p-values} of all test items are showed in Fig.\ref{F3c}, which are all greater than 0.01, indicating that the bit sequence obtained by the experiment has superior performance at random.

\section{\label{sec:5}Discussion}
In this section, we describe the influence of circuit parameters on QRNG randomness, and analyze some attractive properties of the Boson sampling-based QRNG.
\subsection{\label{sec:5.1}Influence of Boson sampling circuit parameters on randomness}

\begin{figure*}[tb]
\begin{center}
\subfigure[]{ \includegraphics[width=1.67in]{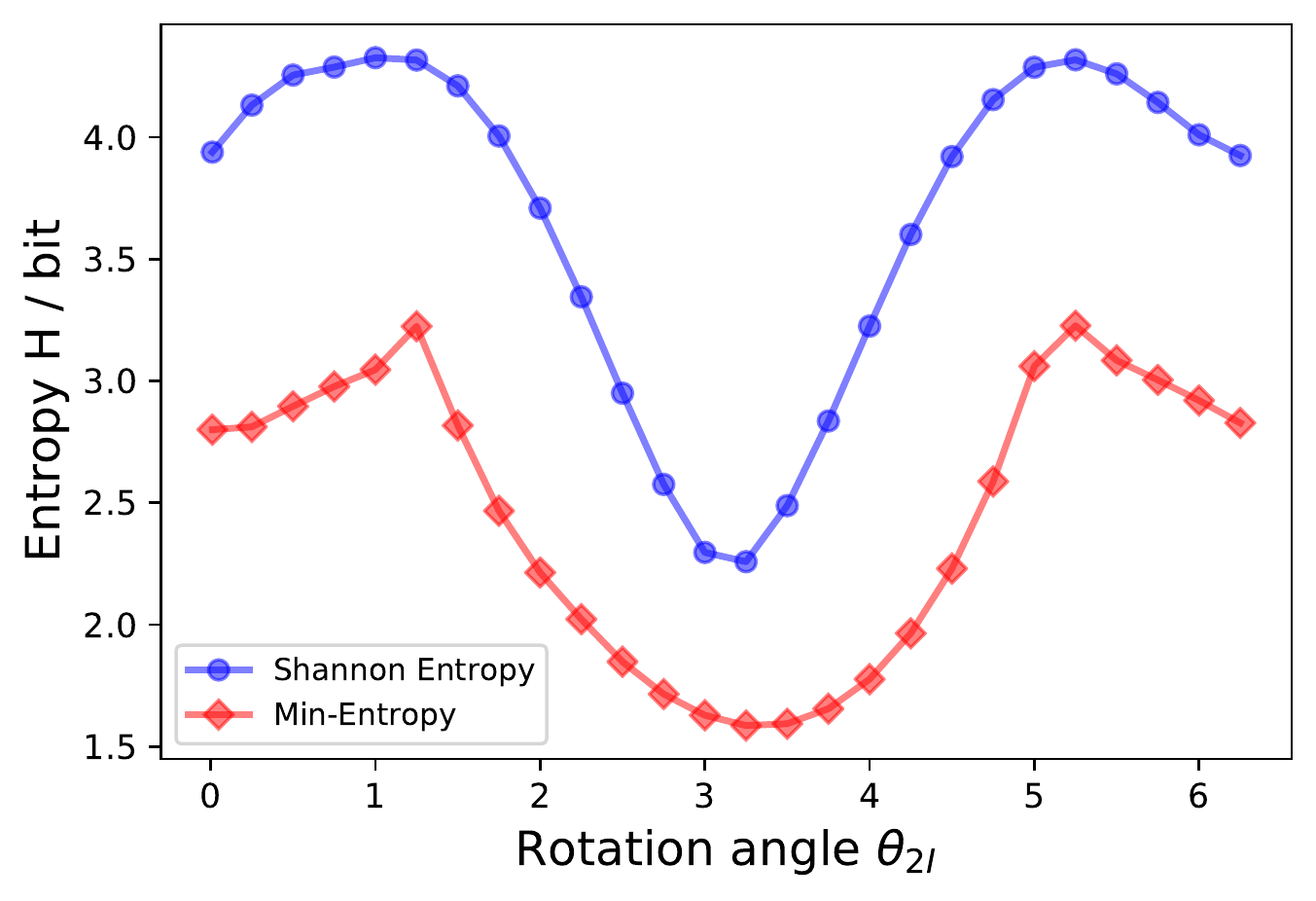}\label{PA1}}
\subfigure[]{ \includegraphics[width=1.67in]{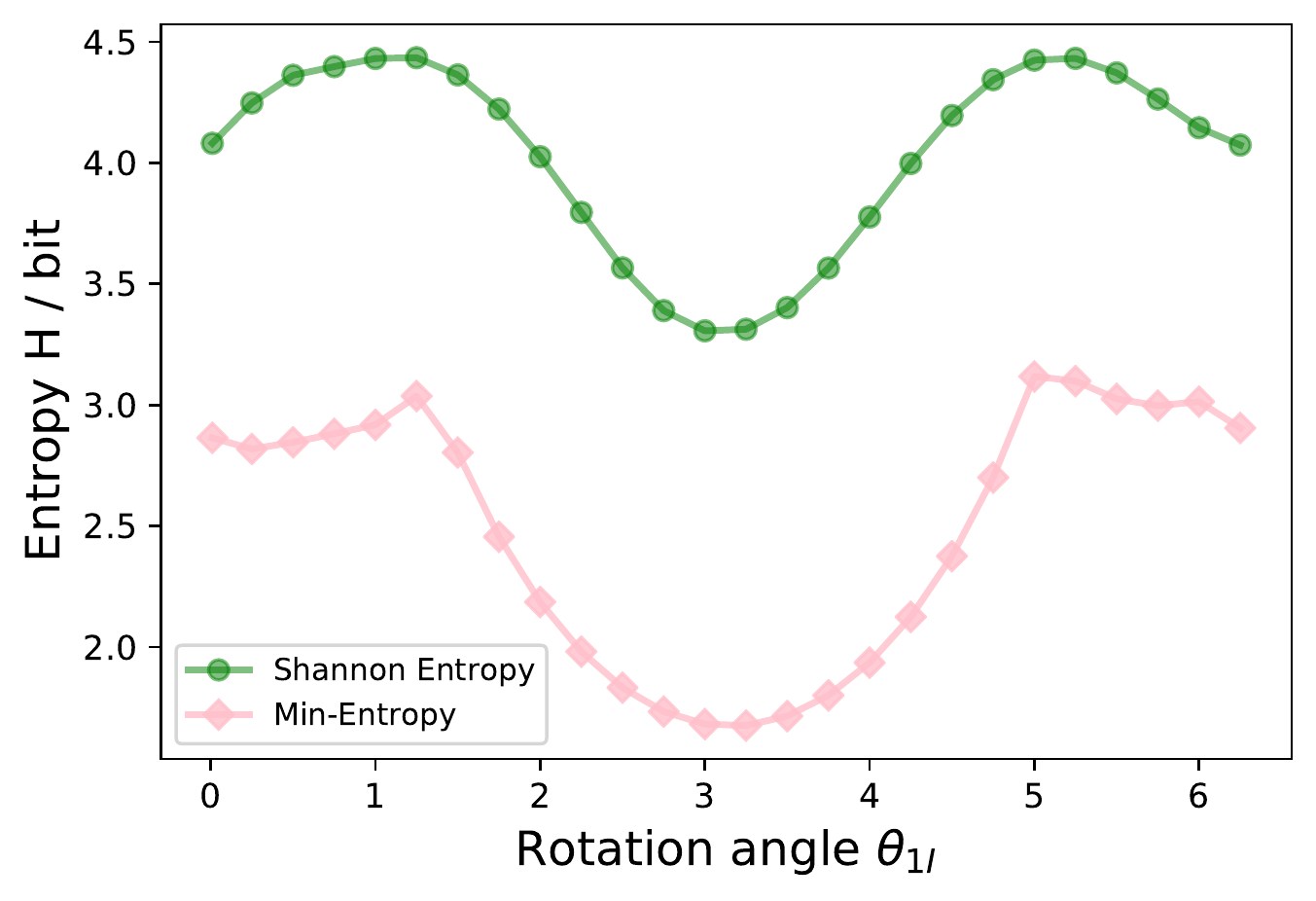}\label{PA2}}
\subfigure[]{ \includegraphics[width=1.67in]{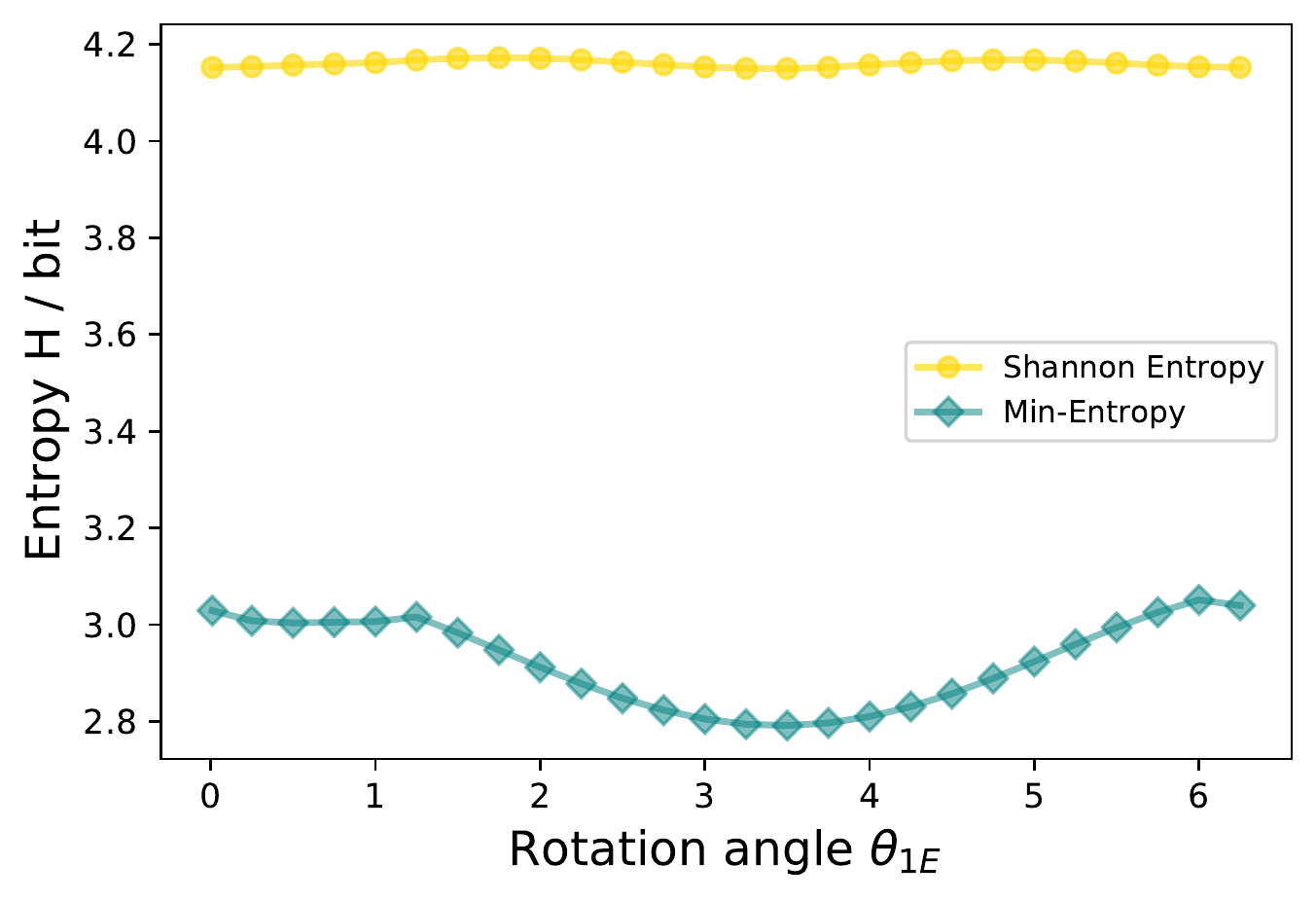}\label{PA3}}
\subfigure[]{ \includegraphics[width=1.67in]{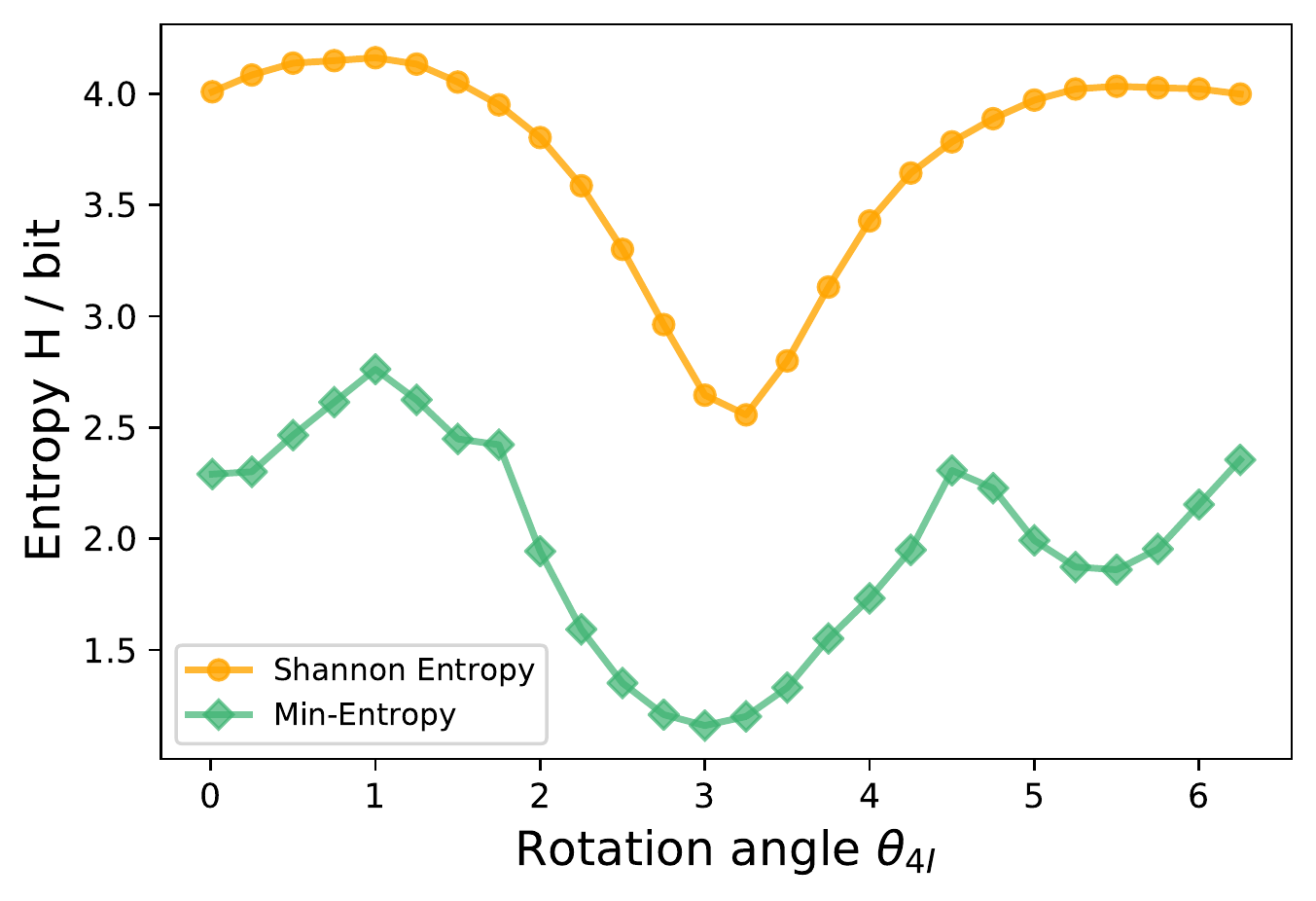}\label{PA4}}
\subfigure[]{ \includegraphics[width=1.67in]{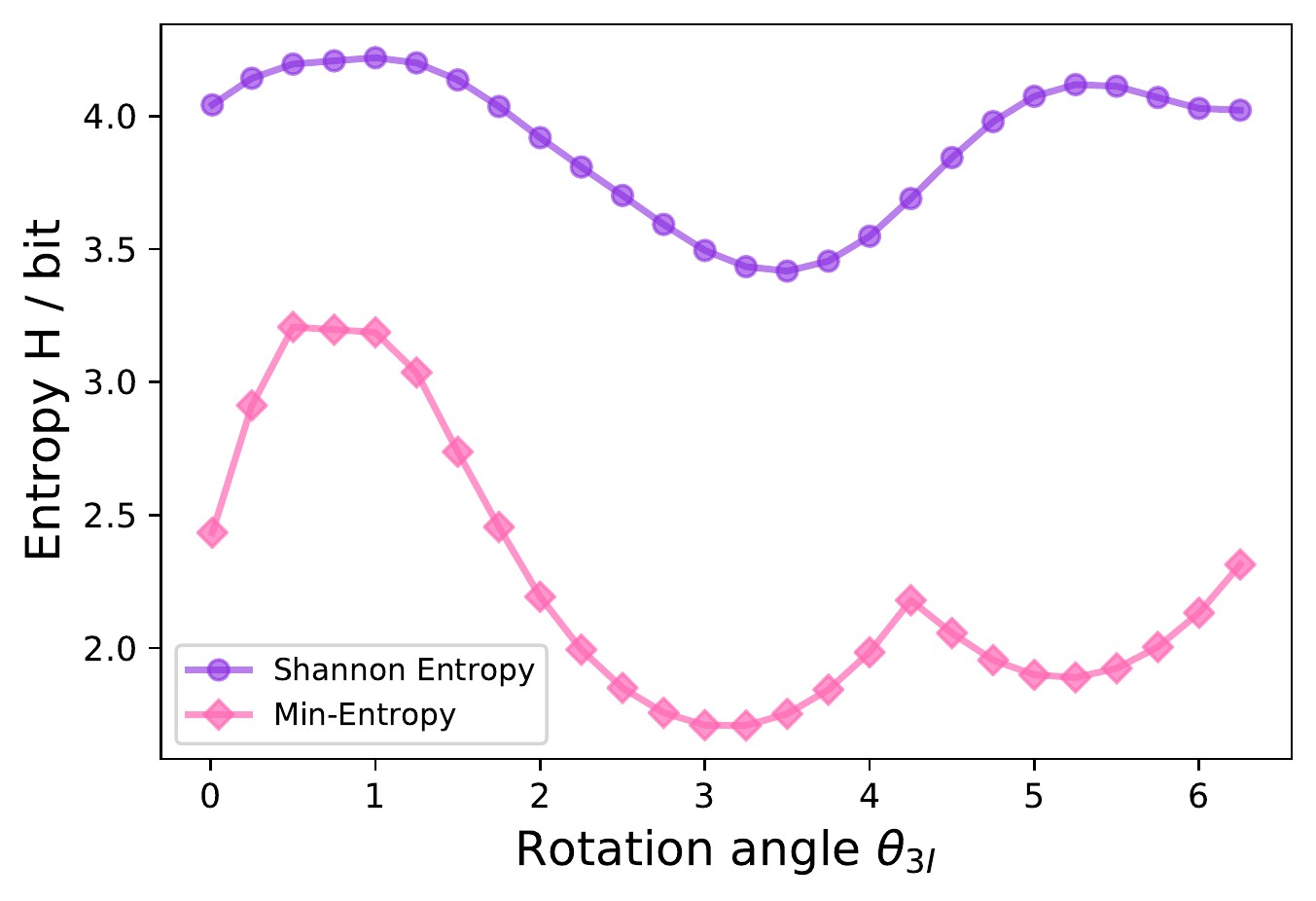}\label{PA5}}
\subfigure[]{ \includegraphics[width=1.67in]{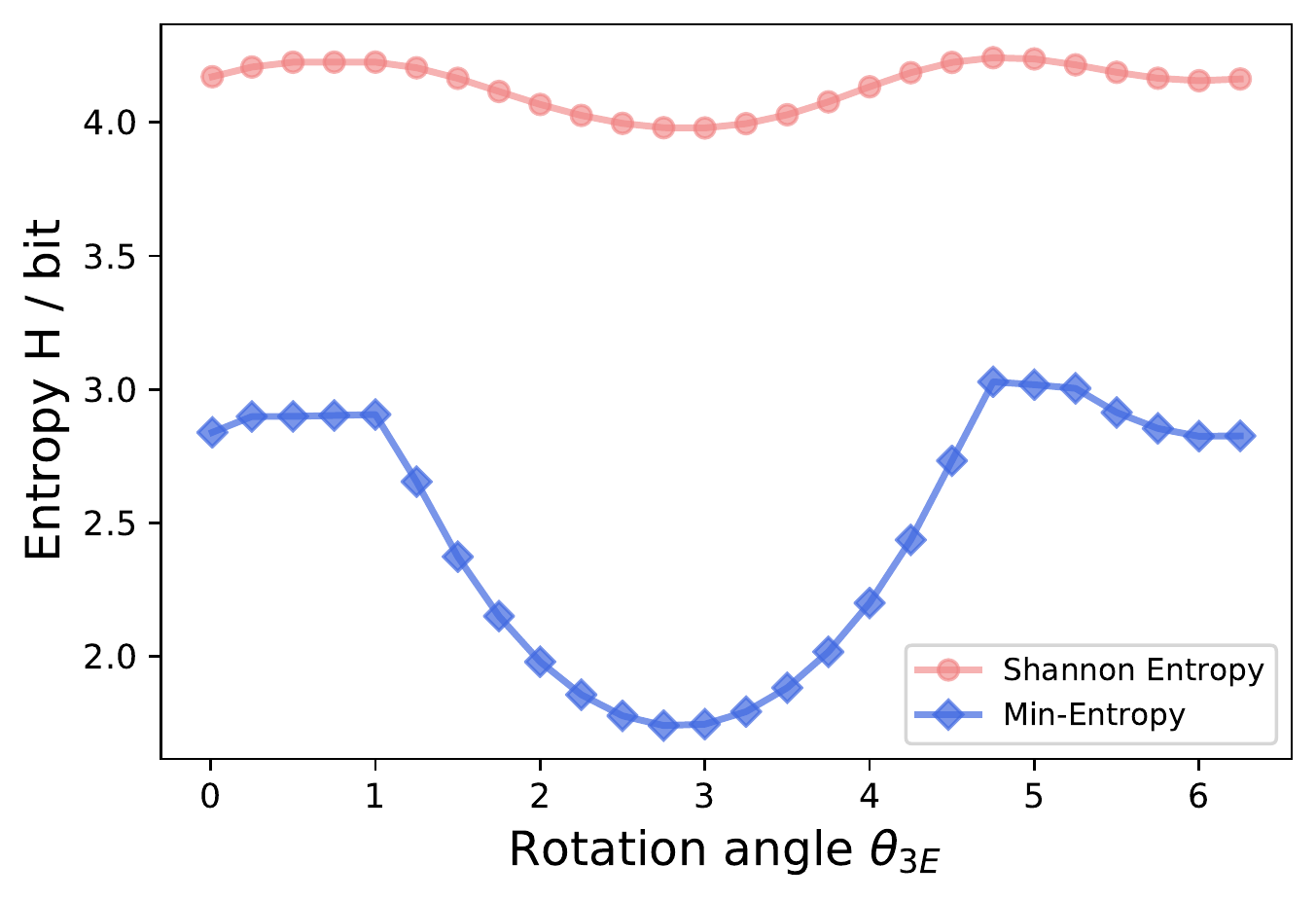}\label{PA6}}
\subfigure[]{ \includegraphics[width=1.67in]{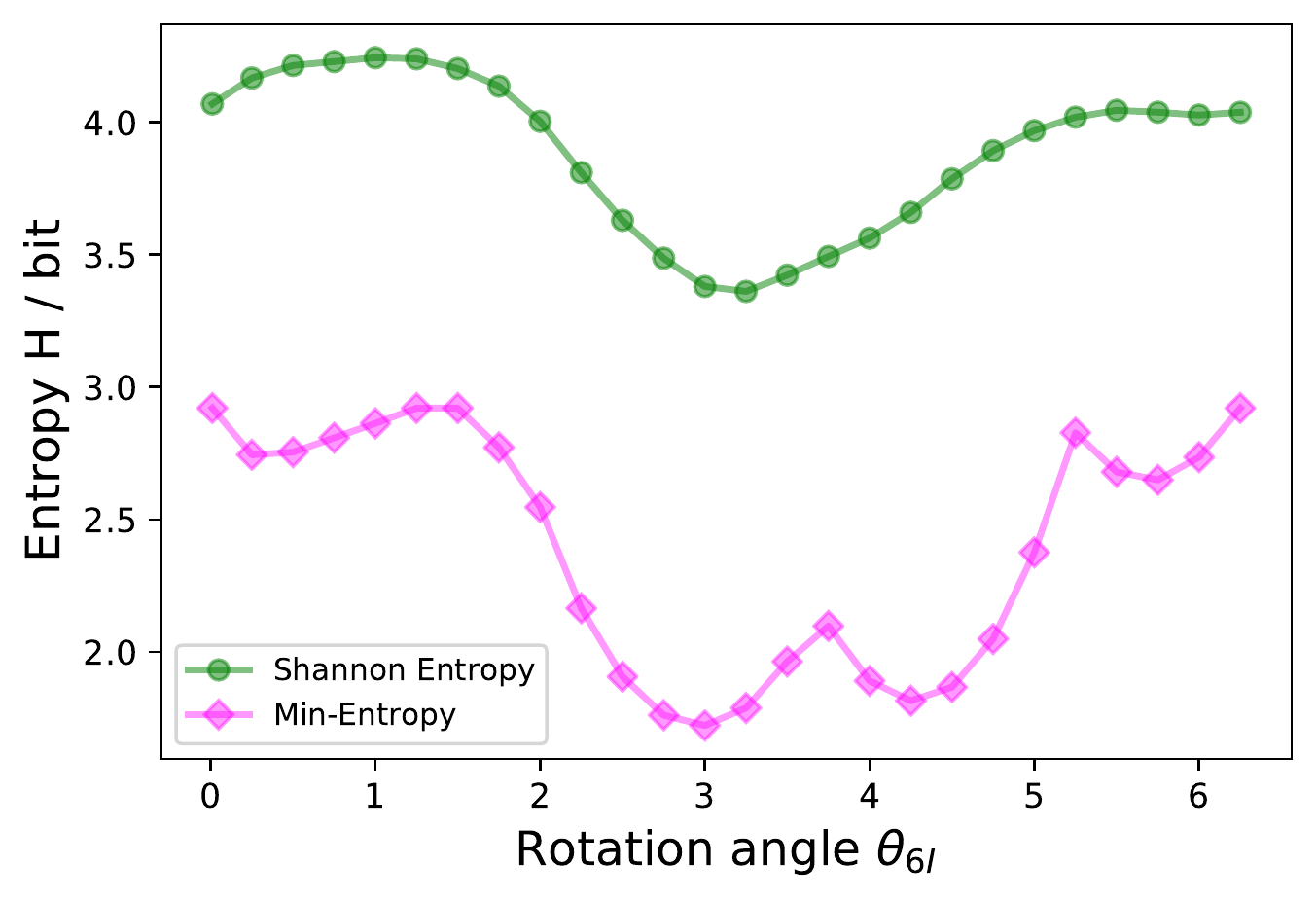}\label{PA7}}
\subfigure[]{ \includegraphics[width=1.67in]{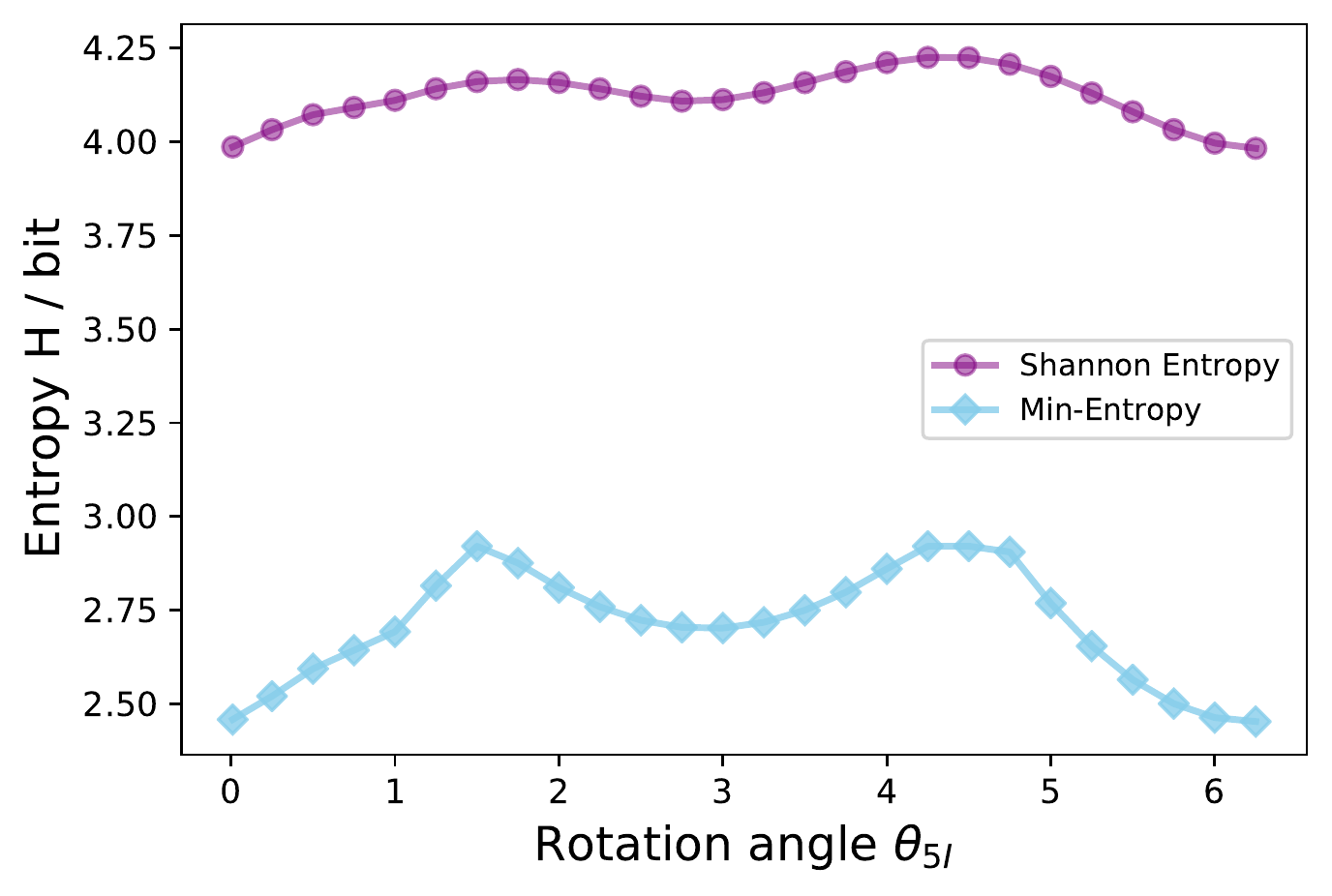}\label{PA8}}
\subfigure[]{ \includegraphics[width=1.67in]{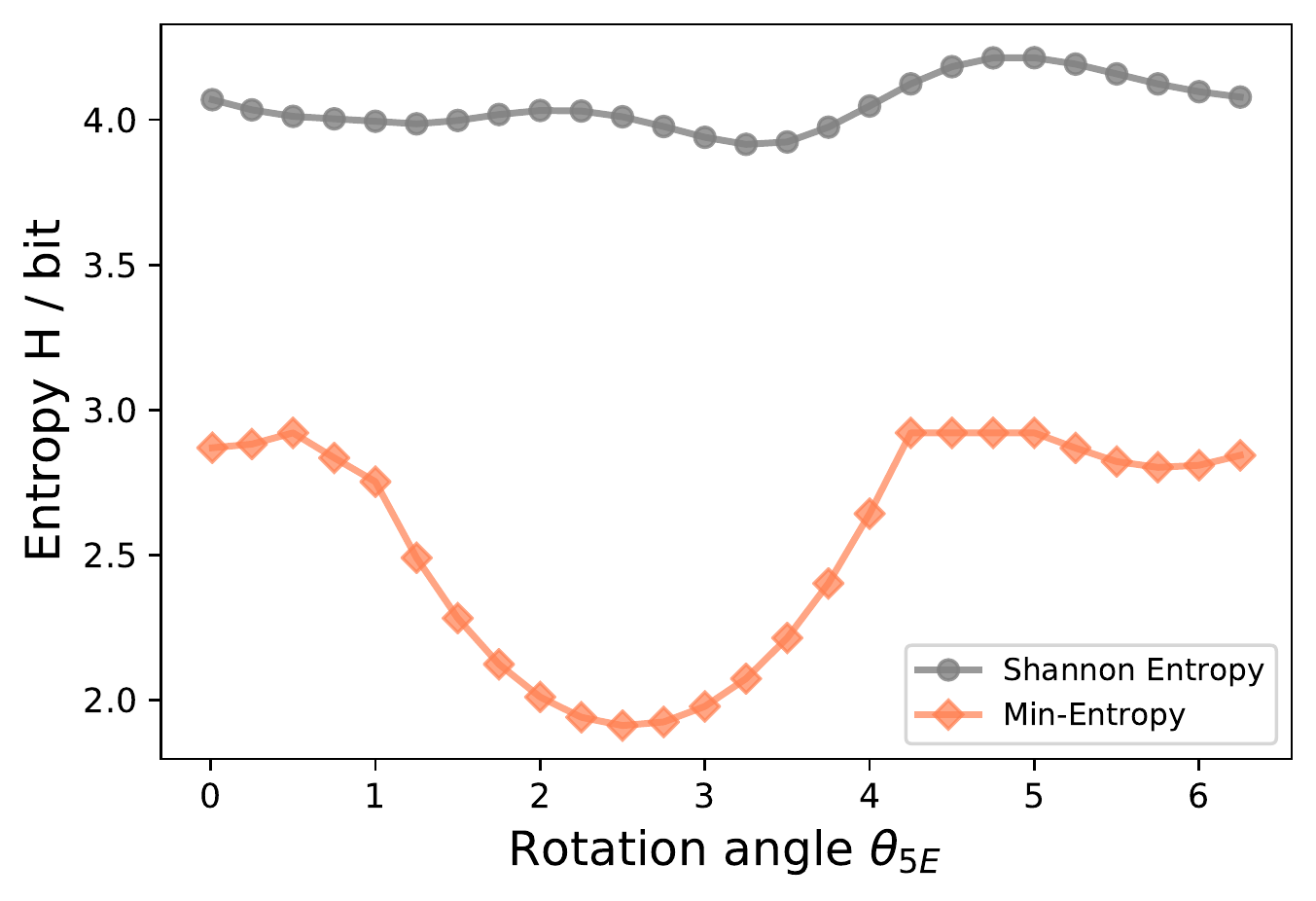}\label{PA9}}
\subfigure[]{ \includegraphics[width=1.67in]{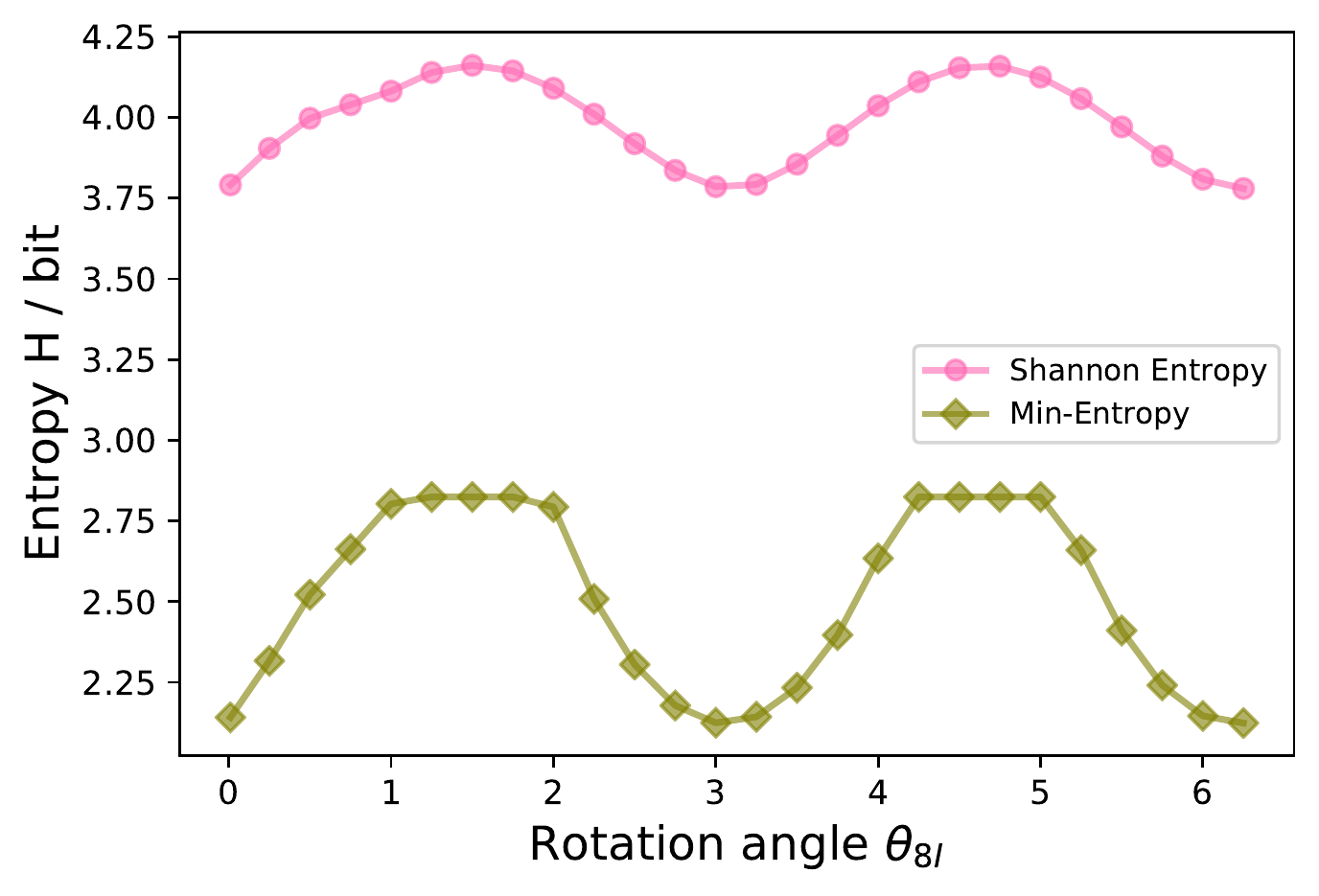}\label{PA10}}
\end{center}
\caption{(a)-(j) present the curves of Shannon entropy and Min-entropy changing with circuit rotation angle parameters $\theta_{2I}$, $\theta_{1I}$, $\theta_{1E}$, $\theta_{4I}$, $\theta_{3I}$, $\theta_{3E}$, $\theta_{6I}$, $\theta_{5I}$, $\theta_{5E}$, $\theta_{8I}$, respectively.}
\label{PAAN}
\end{figure*}

\begin{figure*}[tb]
\begin{center}
\subfigure[]{ \includegraphics[width=3.2in]{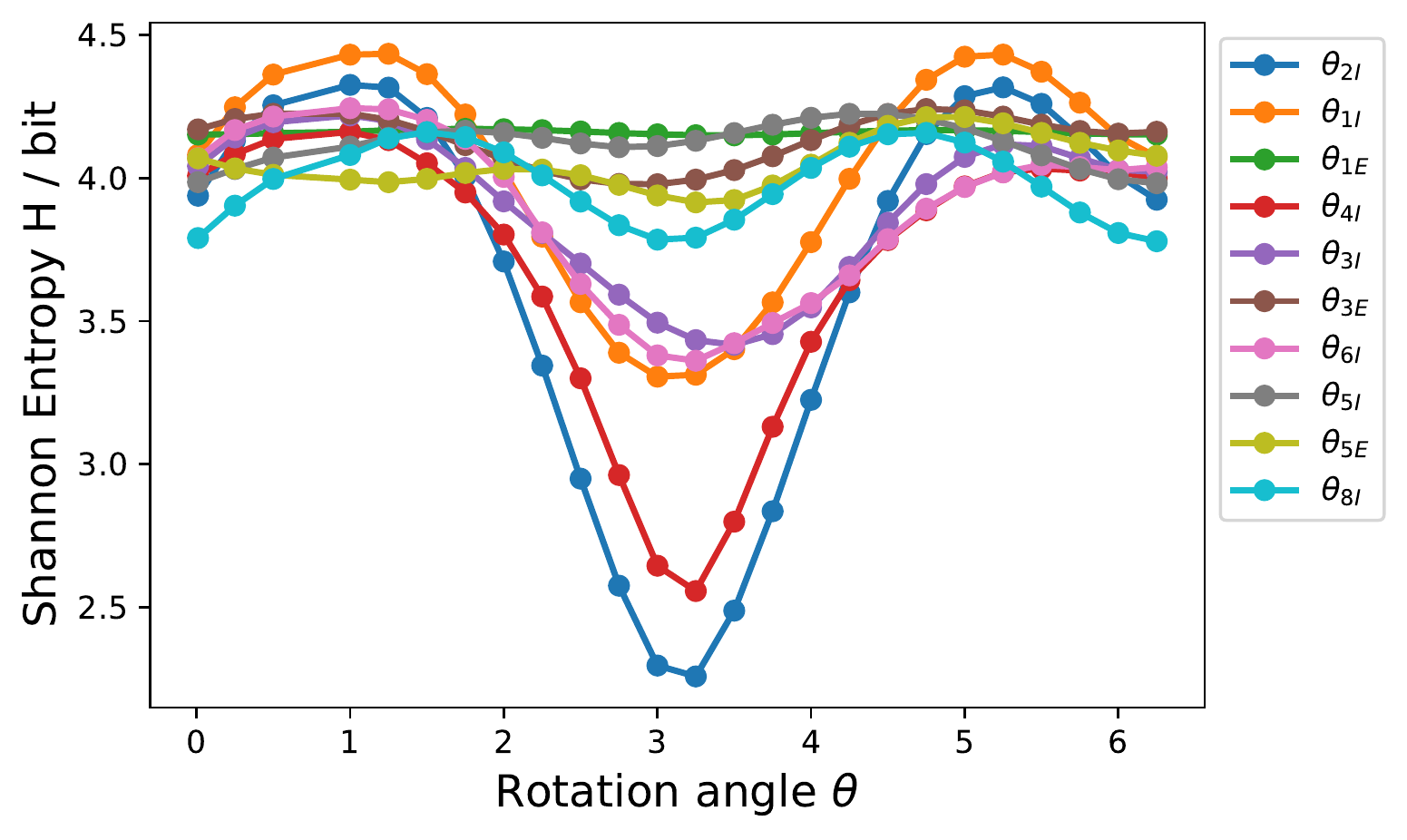}\label{PA11}}
\subfigure[]{ \includegraphics[width=3.2in]{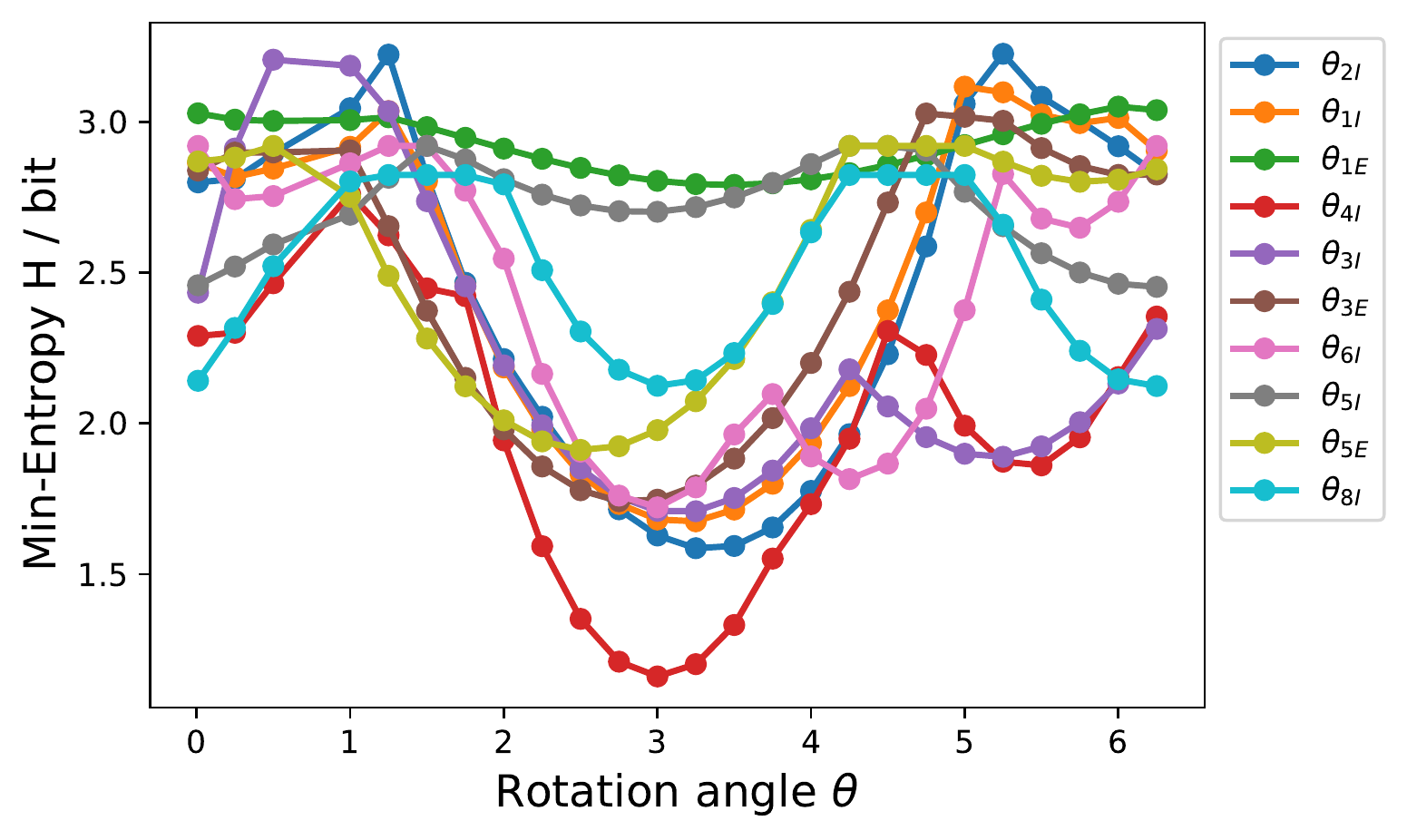}\label{PA12}}
\end{center}
\caption{(a) denotes the curves of Shannon entropy changing with circuit rotation angle parameters.
}
\label{PAAN2}
\end{figure*}

We use mathematical entropy function to measure the uncertainty and randomness of the Boson sampling circuit output in  the information theory, where different entropy values reflect the degree of surprise for the result.
Shannon entropy is calculated as
$H(A)=-\sum_{a \in \chi} P_{A}(a) \log _{2} P_{A}(a)$,
where $ P_{A}$  is the probability distribution of random variable $A$, $P_{A}(a)$  is the probability of getting output $a$, and $\chi$ is a discrete set of values of random variable $A$. \'Rnyi entropy is an extension of the usual Shannon entropy \cite{renyi1961measures}. The \'Rnyi entropy of order $\beta$ is calculated as
$H_{\beta}(A)=\frac{1}{1-\beta} \log _{2} \sum_{a \in \chi} P_{A}(a)^{\beta}$. When $\beta \rightarrow 1 $, the \'Rnyi entropy corresponds to Shannon entropy. When $\beta \rightarrow \infty $, the \'Rnyi entropy corresponds to the Min-entropy. In another way, the Min-entropy $H_{\infty}$ can be defined as
$
H_{\infty}=-\log _{2}\left[\max _{a \in \chi} P_{A}(a)\right]$.
$2^{-H_{\infty}(A)}$ describes the probability of hitting the first guess from a random variable $A$ with a known probability distribution.

In the case of fixed input Boson sampling, different linear interferometer unitary matrices are constructed to obtain the output probability distribution of Boson sampling by adjusting the rotation parameters of the linear optical network phase shifters. As shown in Fig.\ref{PAAN} and Fig.\ref{PAAN2}, the Shannon entropy and Min-entropy of the Boson sampling output distribution vary with different circuit parameters in the range of $\left[0,2 \pi\right)$. The parameters of the 1I and 2I interferometers have a greater impact on the randomness of the circuit output probability distribution, which are the first optical device that the photons entering the optical network from the light source needed to pass. The general trend of each entropy curve is ascending, descending, ascending, descending and repeating.
The experimental results show that for a fixed input 5 modes 2 photons Boson sampling, the randomness of the output of the Boson sampling system can be effectively increased by selecting appropriate linear optical network parameters to improve the random numbers generation rate of Boson sampling-based QRNG.

\subsection{\label{sec:5.2}Performance analysis}

The output probability distribution of Boson sampling is not close to the uniform distribution in variation distance \cite{aaronson2014bosonsampling}, which is the advantage of Boson sampling compared with other sampling distributions. A uniform distribution of 01 bits in the output sequence is an important feature required by the random number generator, where the probability of each bit being 0 or 1 is both 0.5, which is unbiased (otherwise the possibility of "guessing" successfully increases). The Von Neumann correction method is used to achieve unbiasedness in the proposed scheme. However, we have not changed the output probability distribution generated by Boson sampling, which is still far from the uniform distribution. After finishing the post-processing, the probability of each mode coded as 0 and 1 is equivalent, the 01 bits in the final random number sequence are also a uniform distribution at the same time. Different from other schemes that directly change the probability distribution of quantum systems, this scheme has the advantage that it retains the non-uniform distribution of boson sampling. In addition, the Boson sampling-based QRNG has several attractive properties.

\noindent
\textsl{\textbf{Unpredictability.}}
The entire system is in a superposition state of $\left|\phi_{\text {out }}\right\rangle=\sum_{i} \lambda_{i}\left|n_{1} n_{2} \ldots n_{n}\right\rangle$ before being detected by photon detectors,
the measurement may destroy the coherence of the superposition state.
According to the Born rule\cite{born2013principles}, when a quantum superposition state is observed and measured, the probability that the state collapses at a specified position is proportional to the square of the coefficient of the state. The measurement result of a quantum system is related to probability, which can never be predicted better than blindly guessing.
At the same time, Boson sampling is a process of sampling in the probability distribution obtained by the evolution of the Boson through the beam splitters and the phase shifters, where the final output result is random and unpredictable.

\begin{figure}[tb]
\begin{center}
\includegraphics[scale=0.45]{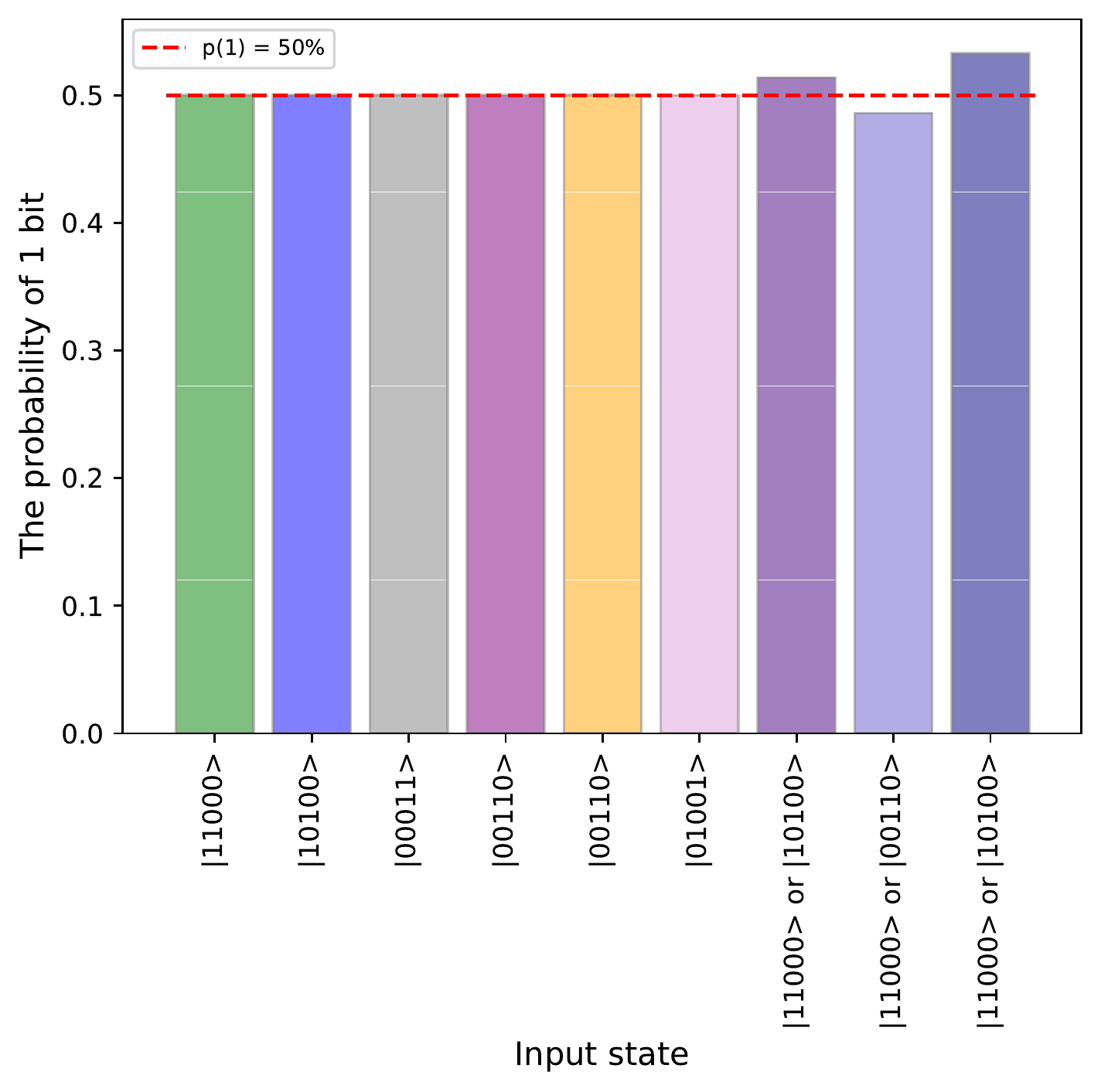}
\end{center}
\caption{After changing different input states, the probability of 1bit in the random number sequence generated by QRNG.}
\label{FIS1}
\end{figure}

\noindent
\textsl{\textbf{Independent of input source.}}
In the traditional QRNGs, the random numbers are generated by three stages: input source, measurement operation, and post-processing, where the original input source of QRNGs must be fully trusted. If the input source is controlled by a third-party, for example,
he/she fixes your source to a combination sequence of ground state $\left|0\right\rangle$ and $\left|1\right\rangle$ instead of the superposition state $\left|+\right\rangle$, when measuring the state on the $Z$ basis, you may "look" to get "random" 0 and 1 bits which are controlled by the third-party.
In the proposed Boson sampling-based QRNG, a haar random linear optical network is placed between the initial input source and the measurement operation,
where the Boson source can be detected and measured by the photon detectors at the output modes after the evolution of the interference.
Since the output state is an unknown superposition state ensured by the interference effect, the result obtained after the measurement can not be completely determined.
In addition, we can also increase the random sampling process before post-processing. Perform boson sampling $T$ times to obtain $T$ sample sets $(T>2)$, from which two sets of samples are randomly selected for post-processing to obtain the output sequence, avoiding the attack problems caused by continuously changing the input source.
We change the input source to other states instead of $\left|11000\right\rangle$ state to simulate the situation that the input source is controlled by the third party. Evolving in the same unitary matrix of the linear optical network, the probabilities of 1 bit in the output random sequence are shown in Figure X, where the obtained random numbers are still unbiased, which are independent of changing input sources.
On the other hand, when the third-party controller alternately gives us different input sources, as shown in Fig.\ref{FIS1}, the probability of generating 1 bit in the random sequence has obvious deviations, which can be easily identified. The random bits will be discarded when there is a significant bias in the 01 bits.
Therefore, regardless of whether the input source is mastered by a third-party, we can generate unbiased random numbers or identify the enemy and discard the generated random numbers, the proposed QRNG is independent of the input source.

\noindent
\textsl{\textbf{Uniform (unbiased).}}
After the Von Neumann correction post-processing, the probability of getting 0 bit and 1 bit in the random number sequences can be equivalent.
For the $x$-th output mode of Boson sampling, the probability of outputting 0 (1) bit is

\begin{widetext}
\begin{equation}
\mathrm{P_x}(0)=\mathrm{P}( detecting\ photon\ in\ S_1) \times \mathrm{P_x}( not\ detecting\ photon\ in\ S_2),
\end{equation}

\begin{equation}
\mathrm{P_x}(1)=\mathrm{P_x}( not\ detecting\ photon\ in\ S_1) \times \mathrm{P_x}( detecting\ photon\ in\ S_2).
\end{equation}
\end{widetext}
In the case of the same input photon source and linear optical network, two consecutive Boson sampling experiments are independent, where $\mathrm{P_x}(detect\ photon\ in\ S_1)=\mathrm{P_x}(detect\ photon\ in\ S_2)$,
so $\mathrm{P_x}(0)=\mathrm{P_x}(1)$.
Therefore, the probability of 0 and 1 in the final random number sequence is both $50\%$, leading to a physically unbiased QRNG in this work.

\noindent
\textsl{\textbf{Unsimulateability.}}
The complexity theory analysis demonstrates that under the plausible conjectures about the permanent of i.i.d. Gaussian matrices\cite{aaronson2013computational}, the problems of accurate and approximate Boson sampling can not be effectively solved by classical computers unless
\begin{equation}\mathrm{PP}=\mathrm{PostBQP}=\mathrm{PostBPP} \subseteq \mathrm{BPP}^{\mathrm{NP}},\end{equation}
which means that the polynomial hierarchy collapses to the third level according to Toda's theorem \cite{toda1991pp}.
The output probability distribution of Boson sampling is calculated by the permanent of the unitary transformation matrix of the linear optical network, which is a $\#P-hard$ task on classical computer. The time complexity of the classic algorithm to calculate the matrix permanent is $ O (2^{n} n^{2})$, where the demand for classical computer resources exponentially increases with $n$ \cite{wu2018benchmark, qiang2016efficient}. Therefore, the Boson sampling-based QRNG can not be simulated classically.

\begin{figure}[tb]
\begin{center}
\subfigure[]{ \includegraphics[width=2.5in]{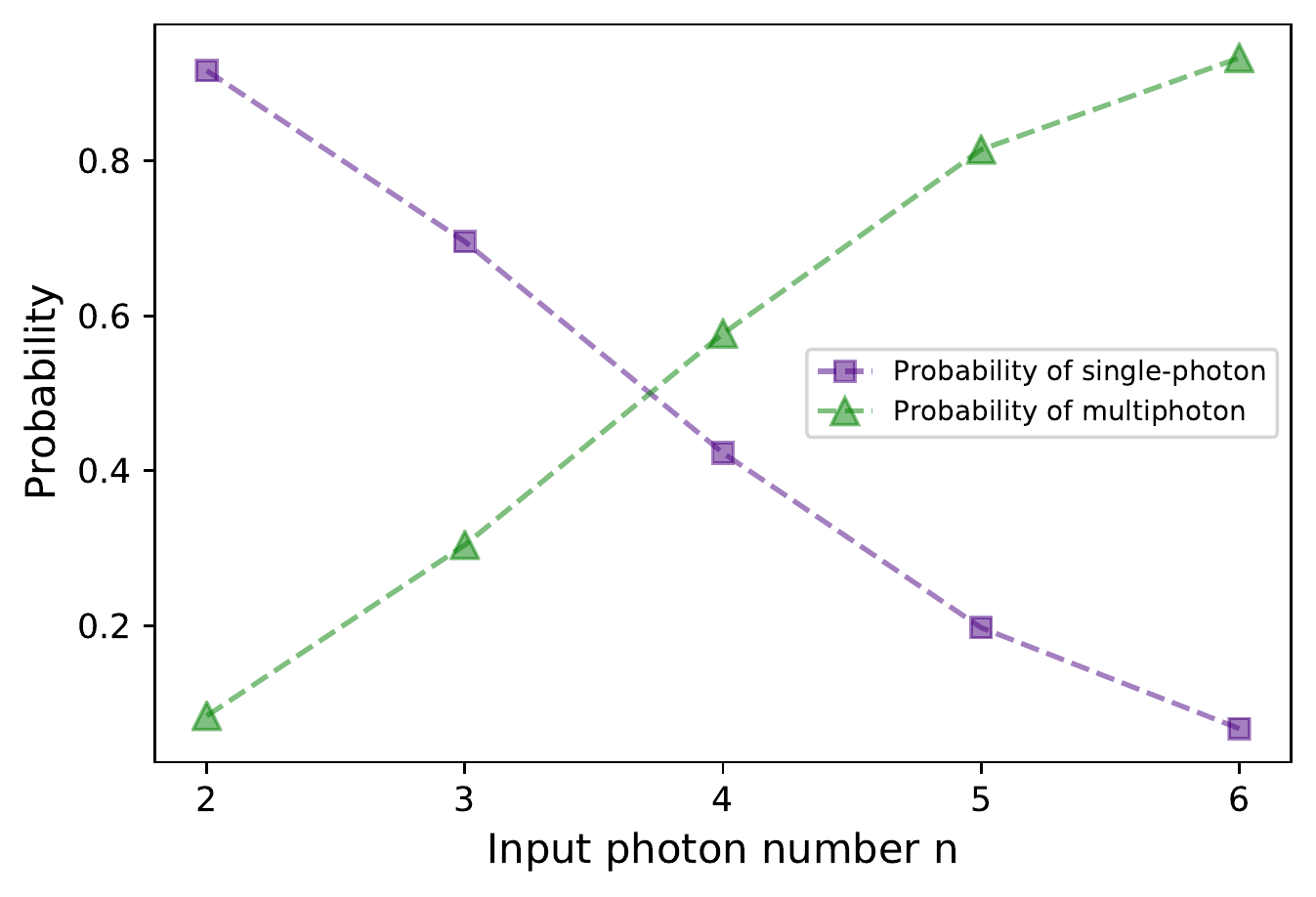}\label{F4b}}
\subfigure[]{ \includegraphics[width=2.51in]{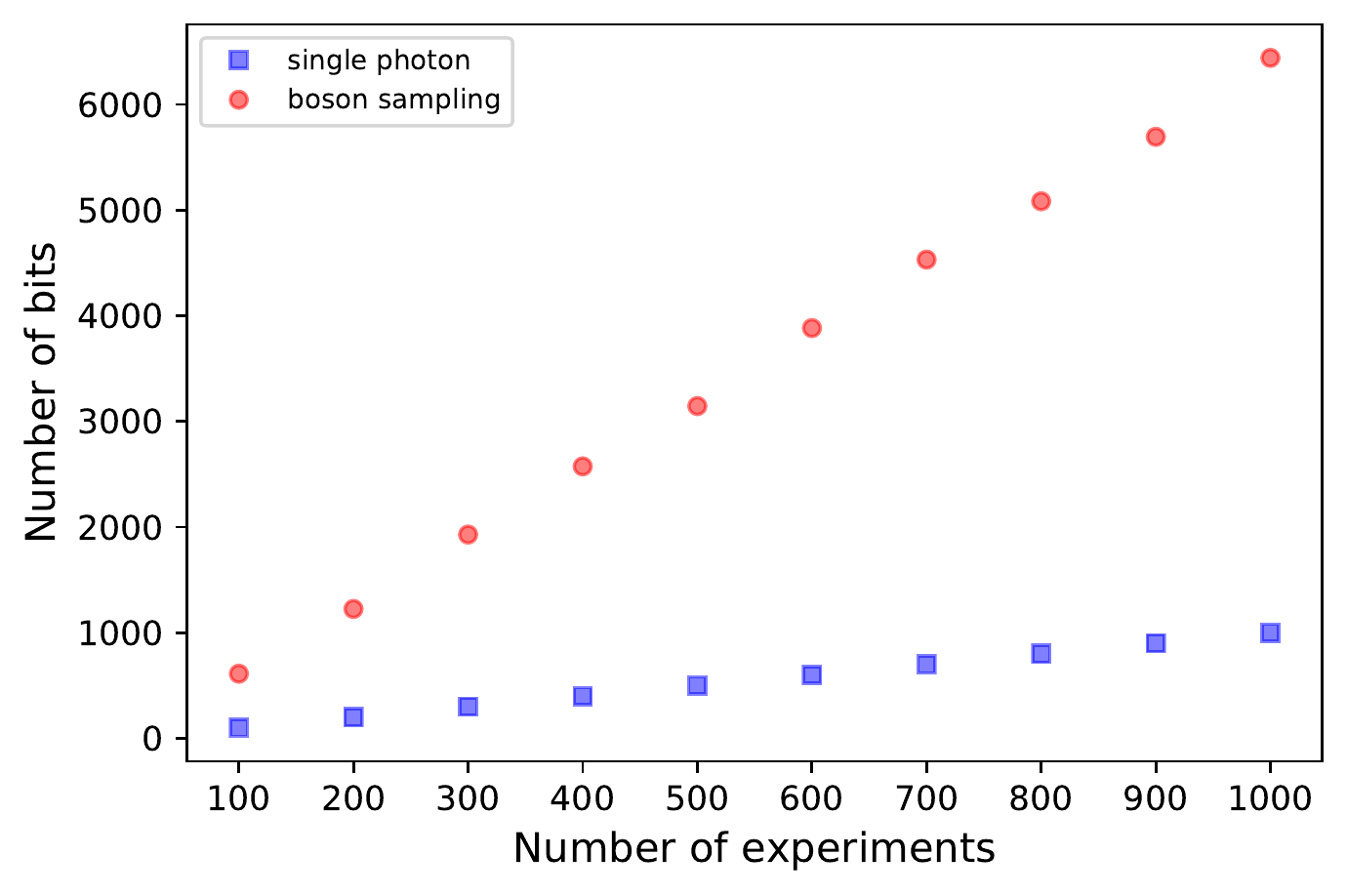}\label{F4c}}
\end{center}
\caption{
(a) In 10 modes Boson sampling with a fixed haar random $U$, the variation curve of the single-photon probability and the multi-photon probability with the number of photons $n$. Under collision-free conditions, the probability of multi-photon in a mode is almost zero. With the increase of the number of photons $n$, the probability of multi-photons gradually increases which cannot be ignored. (b) The number of random number bits generated single photon branching path QRNG and Boson sampling-based QRNG with $M=16, N=6$.
}
\end{figure}

\noindent
\textsl{\textbf{Multiphoton.}}
In the post-processing process, we only record whether a certain mode detects photons instead of the specific number of detected photons.
Compared with the photon counting QRNGs using photon number resolution detectors\cite{ren2011quantum}, the Boson sampling-based QRNG using single-photon detector is relatively easy to realise that it only requires a binary choice between no click (no photon has been detected) and click (one or more photons are detected).
Most of the proposed Boson sampling-based applications rely on collision-free conditions $(n\ll m)$, where the number of photons detected in each output mode is either 0 or 1\cite{aaronson2013computational}, the probability of detecting multiple photons in a mode at the output is almost zero. However, the probability of multiphoton in a mode gradually increases with non-collision-free conditions, seen Fig.\ref{F4b}. The scheme we proposed is not limited to collision-free conditions, where multi-photon and single-photon are handled at the same time in the post-processing, which is more universal.

\noindent
\textsl{\textbf{Multiple random bits.}}
Only one bit can be generated at a time in branching path QRNGs with limited generation rate.
In the scheme mentioned in this article, the linear optical network of Boson sampling has $m$ modes, each of which places a photon detector at the end. Multiple results of the presence or absence of photons can be obtained with one measurement. According to the post-processing principle, more than one random bit can be finally obtained through Boson sampling. We find that the Boson sampling-based QRNG generates random numbers at a faster rate and generates more random numbers at a time by simulating the number of random bits generated by QRNGs of single photon and Boson sampling with $M=16$ and $N=6$ respectively, as shown in Fig.\ref{F4c}.

\section{\label{sec:6}Conclusion}

An available and unbiased quantum random number generator has been proposed by fully exploiting the meaningful randomness of Boson sampling results, in which the sampling result of Boson sampling is encoded to generate binary random number sequences. The experimental results on the Boson sampling-based QRNG prototype system with programmable silicon photonic processor chip passed 15 NIST SP 800-22 statistical test component, indicating that the proposed QRNG scheme performs well in randomness. Performance analysis shows that this scheme can generate random numbers that are uniform, unbiased, unpredictable, source-independent, and non-classical simulation to guarantee security and generates random numbers with multiple bits at a time to speeds up the generation rate simultaneously,
which can solve the problems of the existing discrete QRNGs for the high demand of reliable safe input source, strong photon number resolution and improved bit rate potentially.
In addition, this scheme inherits the merits of Boson sampling, making use of the randomness of Boson sampling results to develop a physical prototype system. This fact provides additional evidence that Boson sampling is capable of solving practical application problems besides quantum advantage. At the same time, it inspires more potential applications of Boson sampling in quantum devices, such as quantum key distribution (QKD) which can be implemented by the synchronization of Boson sampling.

\section*{Acknowledgements}
This work was supported by the National Natural Science
Foundation of China under Grants 61972418 and 61872390, the Natural Science Foundation of Hunan Province
under Grant 2020JJ4750, the Special Foundation for Distinguished
Young Scientists of Changsha under Grant kq2106014, the CCF-Baidu Open Fund under Grants 2021PP15002000, and the Open Fund of Advanced Cryptography and System Security Key Laboratory of Sichuan Province under Grants SKLACSS-202107.

\bibliographystyle{unsrt}
\bibliography{mybib}






\onecolumn

\end{document}